\documentclass[aps,prd,preprint,groupedaddress]{revtex4-2}
\usepackage[utf8]{inputenc}
\usepackage{slashed}
\usepackage{color}
\usepackage{multirow}
\usepackage{physics}
\usepackage{mathrsfs}
\usepackage[colorlinks=true,citecolor=blue,linkcolor=blue]{hyperref}
\usepackage[english]{babel}
\usepackage[T1]{fontenc}
\usepackage[Lenny]{fncychap}
\usepackage[left=2cm,right=2cm,top=2cm,bottom=2cm]{geometry}
\usepackage{amsmath,amssymb,amsthm,dsfont}
\usepackage{amsfonts}
\usepackage{graphicx}
\usepackage{graphics}
\usepackage{subcaption}
\usepackage{float}
\usepackage{wrapfig}
\usepackage{slashed}
\usepackage{lipsum}
\usepackage{fancyhdr}

\usepackage{dcolumn}
\usepackage{psfrag}
\usepackage{gensymb}
\usepackage{lmodern}
\usepackage{multirow}
\usepackage{subcaption}
\usepackage{epsfig}
\usepackage{booktabs}
\usepackage{amsmath,amsfonts,amssymb}
\usepackage{pstcol,pst-fill,pst-grad}
\usepackage{pstricks,pst-fill,pst-grad}
\usepackage{euscript}
\usepackage{pstricks}
\usepackage{wrapfig}
\usepackage{slashed}
\usepackage[toc,page]{appendix}
\usepackage{here}
\selectfont
\usepackage{babel}

\begin{document}


\title{\textbf {$Z$-boson production via the weak process ${e}^{+}{e}^{-}\rightarrow\mu^{+} \mu^{-}$ in the presence of a circularly polarized laser field}}


\author{M. Ouali, M. Ouhammou, S. Taj and B. Manaut}
\affiliation{Sultan Moulay Slimane University, Polydisciplinary Faculty,
Research Team in Theoretical Physics and Materials (RTTPM), Beni Mellal, 23000, Morocco.}


\date{\today}

\begin{abstract}
In the centre of mass frame, we have studied theoretically the $Z$-boson resonant production in the presence of an intense laser field via the weak process  ${e}^{+}{e}^{-}\rightarrow\mu^{+} \mu^{-}$. Dressing the incident particles by a Circularly Polarized laser field (CP-laser field), at the first step, shows that for a given laser field's parameters, the $Z$- boson cross section decreases by several orders of magnitude. We have compared the the Total Cross Section (TCS) obtained by using the scattering matrix method with that given by the Breit-Wigner approach in the presence of a CP-laser field and the results are found to be very consistent. This result indicates that Breit-Wigner formula is valid not only for the laser-free process but also in the presence of a CP-laser field. The dependence of the laser-assisted differential cross section on the Centre of Mass Energy (CME) for different scattering angles proves that it reaches its maximum  for small and high scattering angles. At the next step and by dressing both incident and scattered particles, we have shown that the CP-laser field largely affects the TCS, especially when its strength reaches $10^{9}\,V.cm^{-1}$. This result confirms that obtained for the elastic electron-proton scattering in the presence of a CP-laser field $[\textit{I. Dahiri et al\,}, \href{https://arxiv.org/pdf/2102.00722.pdf}{\color{blue}{arXiv:2102.00722v1[quant-ph] (2021)}}] $. It is interpreted by the fact that heavy interacting particles require high laser field's intensity to affect the collision's cross section.
\end{abstract}


\maketitle

\section{Introduction}\label{section 1}
\par The field of laser–matter interactions usually deals with the atomic or molecular response to an electromagnetic field \cite{1}. However, due to the enormous technological progress in recent years \cite{2}, it is possible today to produce a high energy by laser radiation, which lies far beyond the typical atomic energy scale. Therefore, lasers not only play an important role in traditional fields like atomic physics or quantum optics, but there are also clear evidences for their application in other fields of physics. In this respect, many theoretical physicists have studied numerous laser-assisted processes in particle  physics and quantum electrodynamics (QED)\cite{3,4,5}.

\par In the framework of electroweak standard model of particle physics \cite{6}, the laser assisted scattering processes could have an important impact on the production and decay processes as it affects the measured quantities that characterize the scattering such as the cross section, the branching ratio and life time. In \cite{7}, the authors prove that the laser radiation has a significant effect on the final product of the $Z$-boson decay. In \cite{8}, the authors shows that the laser field enhances the life time of the pion and affects its modes decay. Similar work has been done for a muon decay in \cite{9}.

\par $Z$-boson\cite{10}, whose mass $(M_{Z} = 91,1876\pm0.0021\,GeV)$ and total decay width $(\Gamma_{Z}=2.4952\pm 0.0023\, GeV)$, is a fundamental particle. Together with the $W$ boson, it is responsible for the weak force, one of the four known fundamental forces that govern the behavior of matter in our universe. It was discovered in 1983 at the proton-antiproton collider SPS (CERN), which was mainly constructed to produce $W$ and $Z$-bosons. This discovery was crowned by Noble Prize in physics to Carlo Rubbia (for the boson) and to Simon van der Meer (for the CERN technology)\cite{11}. It is a massive neutral vector boson mediating the weak interaction. The LEP and SLC were designed mainly to produce the $Z$-boson via the ${e}^{+} {e}^{-}$ annihilation. The LEP accelerator had operated at the centre-of-mass energy, from 1989 to 2000, and until 1995 the running was dedicated to the $Z$-boson energy region. From 1996 to 2000, the CME was increased to 161 GeV in order to allow the $W$ and $Z$-boson pair production \cite{12}.  Future linear (ILC ; CLIC) and circular (FCC ; CEPC) electron-positron colliders offer a rich physics program to test the standard model with a high precision measurement and search for new physics beyond the standard model. One of the aims of  CEPC project is  to operate as a super $Z$ factory to create one trillion $Z$-bosons. The $Z$-boson factory allows the production of a vast amount of quarks and leptons via $Z$-boson decay.

In our recent work \cite{13}, we have shown that the CP-laser field decreases the TCS of the Higgs strahlung boson production. Moreover, muon pair production is studied in Quantum electrodynamics (QED) via electron positron annihilation (${e}^{+} {e}^{-} \rightarrow \gamma \rightarrow {\mu}^{+} {\mu}^{-}$) in the presence of both circularly \cite{14} and linearly \cite{15} polarized laser field. It is found that the insertion of a CP-laser field decreases the cross section, while the linearly polarized laser field enhances it. In this respect, we have investigated the process ${e}^{+} {e}^{-} \rightarrow {\mu}^{+} {\mu}^{-}$ at the lowest order, via the $Z$-boson exchange, in the presence a circularly polarized electromagnetic field. It is well known that the $(\gamma, Z)$ interference dominates at the off-peak energies, but its leading contribution to the scattering process  (${e}^{+} {e}^{-} \rightarrow {\mu}^{+} {\mu}^{-}$) vanishes at the peak where the CME is close to the mass of $Z$-boson ($\sqrt{s}=M_{Z}$). An electron-positron collider running at the peak of $Z$-boson mass is called a $Z$-boson factory. So, muon pair production in such a $Z$ factory is a production via ${e}^{+} {e}^{-}$ annihilation at the $Z$-boson peak.

The aim of this paper is to study the $Z$-boson production as a resonance via ${e}^{+} {e}^{-}$ annihilation process (${e}^{+} {e}^{-} \rightarrow {\mu}^{+} {\mu}^{-}$) in the presence of a strong external laser field. We illustrate the effect of the laser field from the theoretical measurement of the cross section as a function of the CME ($\sqrt{s}$) and the laser field's parameters such as laser's electric field amplitude, its frequency and the number of exchanged photons.

The remainder of this paper is organized as follows: In Sec.\,\ref{section 2}, we start by giving the analytical calculation of the TCS where only the incident particles (${e}^{-}$ and  ${e}^{+}$) are dressed by the laser field. This section is divided into two subsections. The first subsection concerns the calculation of the TCS by using the scattering matrix element. The second is about the TCS given by the Breit-Wigner formula in the presence of a CP-laser field. The results and data found are discussed at the end of this section. We proceed, in Sec.\,\ref{section 3}, with the same manner as in the previous section by dressing both incident and scattered particles by a CP-laser field for both Breit-Wigner formula and the TCS obtained by using the S-matrix element. We show how the CP-laser field affects the total TCS by comparing the results found with those obtained in Sec.\,\ref{section 2}. In this work natural units $(\hbar = c = 1)$ are adopted, the choice made for Livi-Civita tensor is that $\varepsilon^{0123}=1$ and the metric $g^{\mu\nu}$ is chosen such that $g^{\mu\nu}=(1,-1,-1,-1)$. 
\section{Free muon production via $Z$-boson exchange in the presence of a CP-laser field} \label{section 2}
In this part, we present the theoretical calculation of the TCS of the scattering process ${e}^{+}{e}^{-}\rightarrow \mu^{+}\mu^{-} $ by only dressing  the incident particles by a CP-laser field. We consider the TCS obtained by both Breit-Wigner formula and that obtained by using the scattering matrix theory. The pair of muon produced are considered free, which means that the laser field only interacts with both electron and positron. The obtained results are discussed at the end of this section.  
\subsection{Theoretical framework}
\subsubsection{Calculation of the total cross section by using the S-matrix}\label{subsection 1}
We consider the weak interaction process ${e}^{+} {e}^{-} \rightarrow {\mu}^{+} {\mu}^{-}$through the $Z$-boson exchange 
\begin{figure}[H]
  \centering
      \includegraphics[scale=0.4]{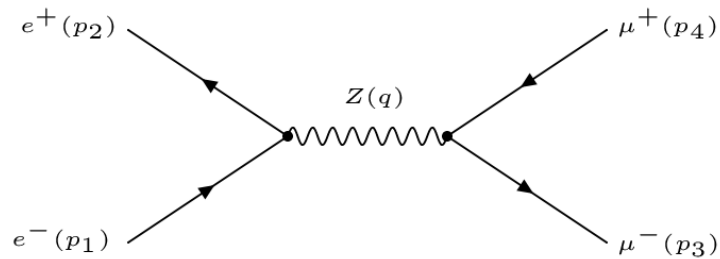}
  \caption{Feynman diagram  for the $s$-channel muon pair production via $e^{+}e^{-}$ collision (annihilation) in the lowest order through the $Z$-boson exchange.}
\end{figure}
The laser field is taken to be monochromatic and circularly polarized. Its classical four vector potential is defined as:
\begin{equation}
A^{\mu}(\phi)=a_{1}^{\mu}\cos\phi+a_{2}^{\mu}\sin\phi \hspace*{1cm};\hspace*{1cm} \phi=(k.x),
\label{1}
\end{equation}
where $a_{1,2}^{\mu}$ are four vectors chosen such as  $a_{1}^{\mu}=(0,a,0,0)$ and $a_{2}^{\mu}=(0,0,a,0)$ with $a$ denotes the amplitude of the four-vector potential. They are orthogonals, which implies that $(a_{1}.a_{2})=0$ and $a_{1}^{2}=a_{2}^{2}=a^{2}=\varepsilon_{0}^{2}/\omega$ with $\varepsilon_{0}$ is the laser's electric field amplitude. $\vec{k}$ is the electromagnetic wave vector with $k^{2}=0$, $\phi$ is the phase of the laser field and $\omega$ its frequency. The Lorentz gauge is satisfied $\partial_{\mu}A^{\mu}=0$, which implies that $k_{\mu}A_{\mu}=0$. The vector  $\vec{k}$ is taken to be parallel to the $z$-axis ($k=(\omega,0,0,\omega)$).
The scattering matrix element \cite{16} for the $Z$-boson production process   ${e}^{+} {e}^{-} \rightarrow {\mu}^{+} {\mu}^{-}$ can be written as: 
\small
\begin{equation}
S_{fi}= -i\dfrac{g^{2}}{16 \cos^{2}\theta_{w}} \int d^{4}x \int d^{4}y\, \overline{\psi}_{p_{2},s_{2}}(x)\gamma^{\mu}(g_{v}-g_{a}\gamma^{5})\psi_{p_{1},s_{1}}(x)D_{\mu\nu}(x-y)\overline{\psi}_{p_{3},s_{3}}(y) \gamma^{\nu}(g_{v}-g_{a}\gamma^{5}) \psi_{p_{4},s_{4}}(y),
\label{2}
\end{equation} 
\normalsize
where $x$ and $y$ denote the space-time coordinates of the incidents particles and the produced muon pair , respectively. $ g_{v}=-1+4 \sin^{2}\theta_{w} $ and $ g_{a}=1 $ are the vector and axial vector coupling constants, respectively. $\gamma^{\mu}$ and $\gamma^{\nu}$ are Dirac matrices. The matrix $\gamma^{5}$ is defined as $\gamma^{5}=i\gamma^{0}\gamma^{1}\gamma^{2}\gamma^{3}$.
 $D_{\mu\nu}(x-y)$ is the free $Z$-boson propagator \cite{16}, which is defined as:
\begin{equation}
 D_{\mu\nu}(x-y)=\int \dfrac{d^{4}q}{(2\pi)^4} \frac{e^{-iq(x-y)}}{q^{2}-M_{Z}^{2}} \Bigg[ig_{\mu\nu}-\dfrac{iq_{\mu}q_{\nu}}{M_{Z}^{2}}\Bigg]
 \label{3}
\end{equation}
 $\psi_{p_{1},s_{1}}(x)$ and $\psi_{p_{2},s_{2}}(x)$ are the Dirac-Volkov states \cite{17} of the electron and the positron, respectively. They are expressed by:
\begin{equation}
\begin{cases}
\psi_{p_{1},s_{1}}(x)= \Big[1-\dfrac{e \slashed k \slashed A}{2(k.p_{1})}\Big] \frac{u(p_{1},s_{1})}{\sqrt{2Q_{1}V}} e^{iS(q_{1},s_{1})} \\
\psi_{p_{2},s_{2}}(x)= \Big[1+\dfrac{e \slashed k \slashed A}{2(k.p_{2})}\Big] \frac{v(p_{2},s_{2})}{\sqrt{2Q_{2}V}} e^{iS(q_{2},s_{2})}
\end{cases},
\label{4}
\end{equation}
with
\begin{equation}
\begin{cases}
S(q_{1},s_{1})=-q_{1}x +\frac{e(a_{1}.p_{1})}{k.p_{1}}\sin\phi - \frac{e(a_{2}.p_{1})}{k.p_{1}}\cos\phi\\
S(q_{2},s_{2})=+q_{2}x +\frac{e(a_{1}.p_{2})}{k.p_{2}}\sin\phi - \frac{e(a_{2}.p_{2})}{k.p_{2}}\cos\phi
\end{cases} .
\label{5}
\end{equation} 
 $\psi_{p_{3},s_{3}}(y)$ and $\psi_{p_{4},s_{4}}(y)$ are the free wave functions of the muon and the antimuon such that:
\begin{equation}
\begin{cases}
\psi_{p_{3},s_{3}}(y)= \frac{u(p_{3},s_{3})}{\sqrt{2E_{3}V}} e^{-ip_{3}y}  \\
\psi_{p_{4},s_{4}}(y)= \frac{v(p_{4},s_{4})}{\sqrt{2E_{4}V}} e^{ip_{4}y} 
\end{cases} ,
\label{6}
\end{equation}
 $u(p_{i},s_{i})$ and $v(p_{i},s_{i})(i=1,2,3,4)$ are Dirac bispinors of the charged fermion and antifermion, respectively. $(s_{i})$ denotes its spin states. $q_{i}(i=1,2)=p_{i}+e^{2}a^{2}/2(k.p_{i})$ is the effective momentum of the incident particles in the presence of the laser field such that:
 \begin{equation}
  q_{1}^{2}=q_{2}^{2}=m_{e}^{*^{2}} = (m_{e}^{2}+e^{2}a^{2}),
 \label{7}
 \end{equation}
  where $m_{e}^{*}$ is the effective mass of the electron and positron inside the laser field.
By inserting the Eqs.(\ref{3}), (\ref{4}) and (\ref{6}) in the scattering matrix (Eq.(\ref{2})) and after calculation, we find that the scattering-matrix element becomes:
\small
\begin{equation}
S_{fi}= -i\dfrac{g^{2}}{16 \cos^{2}\theta_{w}}   \dfrac{1}{\sqrt{16Q_{1}Q_{2}E_{3}E_{4}V^{4}}}  \sum_{n=-\infty}^{+\infty} \Bigg[  \dfrac{1}{(q_{1}+q_{2}+nk)^{2}-M_{Z}^{2}}  \Bigg] M_{fi}^{n}\,(2\pi)^{4}\delta^{4}(p_{3}+p_{4}-q_{1}-q_{2}-nk),
\label{8}
\end{equation}
\normalsize
where $n$ is the number of exchanged photons between the laser field and the colliding electron-positron. $M_{fi}^{n}$ is the scattering amplitude, which is given by:
 \begin{equation}
 M_{fi}^{n}=\Big[  \overline{v}(p_{2},s_{2})\Gamma^{n}_{\mu}u(p_{1},s_{1})   \Big]     \Big[   \overline{u}(p_{3},s_{3})\gamma^{\mu}(g_{v}-g_{a}\gamma^{5})v(p_{4},s_{4}) \Big],
 \label{9}
 \end{equation}
with
\begin{equation}
\Gamma^{n}_{\mu}=M^{0}_{\mu}\,\zeta_{0n}(z)+M^{1}_{\mu}\,\zeta_{1n}(z)+M^{2}_{\mu}\,\zeta_{2n}(z).
\label{10}
\end{equation}
The coefficients $\zeta_{0n}(z$), $\zeta_{1n}(z)$ and $\zeta_{2n}(z)$ are expressed in terms of ordinary Bessel functions as follows:
\begin{equation}
 \left.
  \begin{cases}
     \zeta_{0n}(z) \\
      \zeta_{1n}(z) \\
      \zeta_{2n}(z)
  \end{cases}
  \right\} = \left.
  \begin{cases}
     J_{n}(z)e^{in\phi _{0}}\\
    \frac{1}{2}\big(J_{n+1}(z)e^{i(n+1)\phi _{0}}+J_{n-1}(z)e^{i(n-1)\phi _{0}}\big) \\
     \frac{1}{2i}\big(J_{n+1}(z)e^{i(n+1)\phi _{0}}-J_{n-1}(z)e^{i(n-1)\phi _{0}}\big)
  \end{cases}
  \right\} ,
  \label{11}
\end{equation}
the Bessel function argument is expressed by: 
$z=\sqrt{\alpha_{1}^{2}+\alpha_{2}^{2}}$ and  $\phi_{0}= \arctan(\alpha_{2}/\alpha_{1})$, where
\begin{center}
$\alpha_{1}=\dfrac{e(a_{1}.p_{1})}{(k.p_{1})}-\dfrac{e(a_{1}.p_{2})}{(k.p_{2})}$  \hspace*{1cm};\hspace*{1cm} $\alpha_{2}=\dfrac{e(a_{2}.p_{1})}{(k.p_{1})}-\dfrac{e(a_{2}.p_{2})}{(k.p_{2})}$.\\
\end{center}
The quantities $M^{0}_{\mu}$, $M^{1}_{\mu}$ and $M^{2}_{\mu}$ that appear in the Eq.(\ref{10}) are expressed by:
\begin{equation}
\begin{cases}M^{0}_{\mu}=\gamma_{\mu}(g_{v}-g_{a}\gamma^{5})+2c_{p_{1}}c_{p_{2}}a^{2}k_{\mu}\slashed k(g_{v}-g_{a}\gamma^{5})   &\\
M^{1}_{\mu}=c_{p_{1}}\gamma_{\mu}(g_{v}-g_{a}\gamma^{5})\slashed k\slashed a_{1}-c_{p_{2}}\slashed a_{1}\slashed k \gamma_{\mu}(g_{v}-g_{a}\gamma^{5})   &\\
M^{2}_{\mu}=c_{p_{1}}\gamma_{\mu}(g_{v}-g_{a}\gamma^{5})\slashed k\slashed a_{2}-c_{p_{2}}\slashed a_{2}\slashed k \gamma_{\mu}(g_{v}-g_{a}\gamma^{5}) \end{cases} ,
\label{12}
\end{equation}
with $c_{p_{i}}(i=1, 2)=e/2(k.p_{i})$.
In general, the cross section is calculated by dividing the squared scattering-matrix element (Eq.\ref{8}) by $VT$ and by the incoming particles' current $|J_{in}|=\sqrt{(q_{1}q_{2})^{2}-m_{e}^{*^{2}}}/Q_{1}Q_{2}V$ \cite{16}. Then we multiply the result by the density of final states. By summing over the final states and averaging over the initial states, we get the partial differential cross section:
\small
\begin{multline}
\dfrac{d\sigma_{n}}{d\Omega}=  \dfrac{g^{4}}{256\times 4\cos^{4}\theta_{w}} \dfrac{1}{\sqrt{(q_{1}q_{2})^{2}-m_{e}^{*^{4}}}} \sum_{n=-\infty}^{+\infty} \Big[  \dfrac{1}{\big[(q_{1}+q_{2}+nk)^{2}-M_{Z}^{2}\big]^{2}+M_{Z}^{2}\Gamma_{Z}^{2}} \Big] |\overline{M_{fi}^{n}}|^{2} \\   \int \dfrac{d^{3}p_{3}}{(2\pi)^{3}E_{3}} \int \dfrac{d^{3}p_{4}}{(2\pi)^{3}E_{4}}  (2\pi)^{4}\delta^{4}(p_{3}+p_{4}-q_{1}-q_{2}-nk),
\label{13}
\end{multline}
\normalsize
where
\begin{equation}
 |\overline{M_{fi}^{n}}|^{2}=\frac{1}{4}Tr\big[ (\slashed p_{1}-m_{e})\Gamma^{n}_{\mu}(\slashed p_{2}+m_{e})\overline{\Gamma^{n}_{\nu}}\, \big]  Tr\big[ (\slashed p_{3}-m_{\mu})\gamma^{\mu}(g_{v}-g_{a}\gamma^{5}) (\slashed p_{4}+m_{\mu})\gamma^{\nu}(g_{v}-g_{a}\gamma^{5}) \big],
 \label{14}
\end{equation}
with $\Gamma^{n}_{\mu}$ is given by Eq.(\ref{10}),
\begin{equation}
 \overline{\Gamma^{n}_{\nu}}=\overline{M^{0}_{\nu}}\,\zeta_{0n}^{*}(z)+\overline{M^{1}_{\nu}}\,\zeta_{1n}^{*}(z)+\overline{M^{2}_{\nu}}\,\zeta_{2n}^{*}(z) ,
 \label{15}
\end{equation}
and
\begin{equation}
\begin{cases}\overline{M^{0}_{\nu}}=\gamma_{\nu}(g_{v}-g_{a}\gamma^{5})+2c_{p_{1}}c_{p_{2}}a^{2}k_{\nu}\slashed k(g_{v}-g_{a}\gamma^{5})   &\\
\overline{M^{1}_{\nu}}=c_{p_{1}} \slashed a_{1} \slashed k \gamma_{\nu}(g_{v}-g_{a}\gamma^{5}) -c_{p_{2}}\gamma_{\nu}(g_{v}-g_{a}\gamma^{5})\slashed k\slashed a_{1}   &\\   
\overline{M^{2}_{\nu}}=c_{p_{1}} \slashed a_{2} \slashed k \gamma_{\nu}(g_{v}-g_{a}\gamma^{5}) -c_{p_{2}}\gamma_{\nu}(g_{v}-g_{a}\gamma^{5})\slashed k\slashed a_{2} \end{cases}.
\label{16}
\end{equation}
Since the propagator given by Eq.(\ref{3}) doesn't account for the fact that the $Z$-boson is an unstable particle, we have made the following  replacement in Eq.(\ref{13}) to avoid the divergence that appears at $(q_{1}+q_{2}+nk)^{2}=M_{Z}^{2}$
 \begin{equation}
\dfrac{1}{q^{2}-M_{Z}^{2}}\rightarrow \dfrac{1}{q^{2}-M_{Z}^{2}+iM_{z}\Gamma_{Z}}.
\label{17}
\end{equation}
 Then we have:
 \begin{equation}
   \Big|\dfrac{1}{(q_{1}+q_{2}+nk)^{2}-M_{Z}^{2}+iM_{z}\Gamma_{Z}}\Big|^{2}\rightarrow \Bigg| \dfrac{1}{\big((q_{1}+q_{2}+nk)^{2}-  M_{Z}^{2}\big)^{2}+M_{Z}^{2}\Gamma_{Z}^{2}}\Bigg|.
   \label{18}
 \end{equation}
The trace calculation (Eq.$\ref{14}$) is performed by using FEYNCALC program\cite{18}. The expression of $|\overline{M_{fi}^{n}}|^{2}$ can be explicitly expressed in terms of Bessel functions as follows:
\small
\begin{equation}
 |\overline{M_{fi}^{n}}|^{2}=\frac{1}{4}\Big[ AJ_{n}^{2}(z)+BJ_{n+1}^{2}(z)+CJ_{n-1}^{2}(z)+DJ_{n}(z)J_{n+1}(z)+EJ_{n}(z)J_{n-1}(z)+FJ_{n-1}(z)J_{n+1}(z) \Big] .
 \label{19}
\end{equation}
\normalsize
The coefficients $A$, $B$, $C$, $D$, $E$ and $F$ are too long to be presented in this paper. We give only the expression of the first coeffcient multiplied by {$ J_{n}^{2}(z) $}.
\small
\begin{eqnarray}
A&=&\nonumber\dfrac{16}{((k.p_{1}) (k.p_{2}))}\Big[(a^{4} e^{4} (g_{a}^2 + g_{v}^2) (g_{a}^2 + g_{v}^2) (k.p_{3}) (k.p_{4}) + 2 (k.p_{1}) (k.p_{2}) (4 g_{a}^{2} g_{v}^{2} ((p_{1}.p_{4}) (p_{2}.p_{3})\\ &-&\nonumber (p_{1}.p_{3}) (p_{2}.p_{4}))+ g_{v}^2 (g_{a}^2 (-m_{\mu}^{2} (2 m_{e}^{2} + (p_{1}.p_{2})) + (p_{1}.p_{4}) (p_{2}.p_{3}) + (p_{1}.p_{3}) (p_{2}.p_{4}) +m_{e}^{2} (p_{3}.p_{4}))\\&+&\nonumber g_{v}^2 (m_{\mu}^{2} (2 m_{e}^{2} + (p_{1}.p_{2})) + (p_{1}.p_{4}) (p_{2}.p_{3}) + (p_{1}.p_{3}) (p_{2}.p_{4}) + m_{e}^{2} (p_{3}.p_{4})))+g_{a}^2 (g_{a}^2 (-m_{\mu}^{2} (p_{1}.p_{2}) \\&+&\nonumber (p_{1}.p_{4}) (p_{2}.p_{3}) + (p_{1}.p_{3}) (p_{2}.p_{4}) + m_{e}^{2} (2 m_{\mu}^{2} - (p_{3}.p_{4})))+g_{v}^2 (m_{\mu}^{2} (p_{1}.p_{2}) + (p_{1}.p_{4}) (p_{2}.p_{3})\\ &+& (p_{1}.p_{3}) (p_{2}.p_{4})-m_{e}^{2} (2 m_{\mu}^{2} + (p_{3}.p_{4})))))+a^2 e^2 (4 g_{a} g_{a} g_{v} g_{v} ((k.p_{2}) (k.p_{4}) (p_{1}.p_{3}) - (k.p_{2}) (k.p_{3}) \\ & \times &\nonumber(p_{1}.p_{4}) -(k.p_{1}) (k.p_{4}) (p_{2}.p_{3}) + (k.p_{1}) (k.p_{3}) (p_{2}.p_{4})) + g_{a}^2 (g_{a}^2 ((k.p_{2}) (k.p_{4}) (p_{1}.p_{3}) +(k.p_{3})\\ & \times &\nonumber (2 (k.p_{4}) (m_{e}^{2} - (p_{1}.p_{2})) + (k.p_{2}) (p_{1}.p_{4}) + (k.p_{1}) (p_{2}.p_{4}))+(k.p_{1}) (2 (k.p_{2}) m_{\mu}^{2} + (k.p_{4}) (p_{2}.p_{3}) \\ & - &\nonumber 2 (k.p_{2}) (p_{3}.p_{4})))+ g_{v}^2 ((k.p_{2}) (k.p_{4}) (p_{1}.p_{3}) +(k.p_{3}) (2 (k.p_{4}) (m_{e}^{2} - (p_{1}.p_{2})) + (k.p_{2}) (p_{1}.p_{4})\\ & + &\nonumber (k.p_{1}) (p_{2}.p_{4}))+(k.p_{1}) ((k.p_{4}) (p_{2}.p_{3}) - 2 (k.p_{2}) (m_{\mu}^{2} + (p_{3}.p_{4}))))) + g_{v}^2 (g_{a}^2 ((k.p_{2}) (k.p_{4}) (p_{1}.p_{3})\\ & + &\nonumber(k.p_{3}) (-2 (k.p_{4}) (m_{e}^{2} + (p_{1}.p_{2})) + (k.p_{2}) (p_{1}.p_{4}) + (k.p_{1}) (p_{2}.p_{4}))+(k.p_{1}) (2 (k.p_{2}) m_{\mu}^{2} + (k.p_{4})\\ &- &\nonumber (p_{2}.p_{3})  2 (k.p_{2}) (p_{3}.p_{4})))+g_{v}^2 ((k.p_{2}) (k.p_{4}) (p_{1}.p_{3})+(k.p_{3}) (-2 (k.p_{4}) (m_{e}^{2} + (p_{1}.p_{2})) + (k.p_{2})\\ & \times &\nonumber (p_{1}.p_{4})  + (k.p_{1}) (p_{2}.p_{4})) +(k.p_{1}) ((k.p_{4}) (p_{2}.p_{3}) - 2 (k.p_{2}) (m_{\mu}^{2} + (p_{3}.p_{4})))))))\Big].
\label{20}
\end{eqnarray}
\normalsize
To get the differential cross section, we have to integrate the Eq.(\ref{13}) over $d^{3}p_{4}$ and use  $d^{3}p_{3}=|\mathbf{p_{3}}|^{2}d|\mathbf{p_{3}}|d\mathbf{\Omega} $.
\small
\begin{multline}
\dfrac{d\sigma_{n}}{d\Omega}=  \dfrac{g^{4}}{256\times 4\cos^{4}\theta_{w}}\dfrac{1}{\sqrt{(q_{1}q_{2})^{2}-m_{e}^{*^{4}}}} \sum_{n=-\infty}^{+\infty} \Big[  \dfrac{1}{((q_{1}+q_{2}+nk)^{2}-M_{Z}^{2})^{2}+M_{Z}^{2}\Gamma_{Z}^{2} } \Big]
 |\overline{M_{fi}^{n}}|^{2}  \\  \int \dfrac{2|\textbf{p}_{3}|^{2}d|\textbf{p}_{3}|}{(2\pi)^{3}E_{3}}   (2\pi)^{4}\delta^{4}((Q_{1}+Q_{2}+n\omega-E_{3})^{2}-m_{\mu}^{2}).
\label{21}
\end{multline}
\normalsize
The remaining integral over $d|\textbf p_{3}|$ can be evaluated by using the well known formula \cite{16}
\begin{equation}
\int d\textbf x f(\textbf x) \delta(g(\textbf x))=\dfrac{f(\textbf x)}{|g^{'}(\textbf x)|_{g(\textbf x)=0}}.
\label{22}
\end{equation}
 We get:
 \small
\begin{equation}
\dfrac{d\sigma_{n}}{d\Omega}=  \dfrac{g^{4}}{256\times 4\cos^{4}\theta_{w}} \dfrac{1}{\sqrt{(q_{1}q_{2})^{2}-m_{e}^{*^{4}}}} \sum_{n=-\infty}^{+\infty} \Big[  \dfrac{1}{((q_{1}+q_{2}+nk)^{2}-M_{Z}^{2})^{2}+M_{Z}^{2}\Gamma_{Z}^{2} } \Big]|\overline{M_{fi}^{n}}|^{2}   \dfrac{2|\textbf{p}_{3}|^{2}}{(2\pi)^{2}E_{3}} \dfrac{1}{\big|g^{'}(|\textbf{p}_{3}|)\big|},
\label{23}
\end{equation}
\normalsize
where
\begin{equation}
\big|g^{'}(|\textbf{p}_{3}|)\big|=-2\Bigg[\Big[\sqrt{s}+n\omega+\frac{e^{2}a^{2}}{2}\Big(\dfrac{4}{\sqrt{s}}\Big)\Big]\dfrac{|\textbf p_{3}|}{\sqrt{|\textbf p_{3}|^{2}+m_{\mu}^{2}}}\Bigg].
\label{24}
\end{equation}
To reach the well known sum-rule (Kroll Whtason formula) \cite{19}, we have to sum over the cutoff number of exchanged photons. Thus, we get the total differential cross section that corresponds to the laser-free differential cross section
\begin{equation}
\sum_{n=-\infty}^{+\infty}\dfrac{d\sigma_{n}}{d\Omega}=  \Big(\dfrac{d\sigma}{d\Omega}\Big)^{laser-free} .
\label{25}
\end{equation}
The total cross section  might be obtained by numerically integrating over the solid angle $d\Omega=\sin(\theta) d(\theta) d(\phi)$ and summing over the cutoff number of transferred photons
\begin{equation}
\sum_{n=-\infty}^{+\infty}\sigma_{n}= \big(\sigma\big)^{laser-free} .
\label{26}
\end{equation}
\subsubsection{Breit-Wigner resonance for $Z$ production in the presence of a CP-laser field }\label{subsection 2}
In this subsection, we consider the TCS given by the Breit-Wigner formula \cite{20} in the absence of an external field. As long as we are interested in the dressing of the colliding particles with a CP-laser field, we will calculate the laser-assisted partial width of the $Z$ decay into electron and positron to obtain the laser-assisted Breit-Wigner's TCS. 
The TCS of the laser-free process $e^{+}e^{-}\rightarrow \mu^{+} \mu^{-} $ is given in \cite{21} by :
\begin{equation}
\sigma=\frac{1}{192\pi}\dfrac{g^{4}}{\cos^{4}\theta_{w}}\dfrac{M_{Z}}{(s-M_{Z}^{2})^{2}+M_{Z}^{2}\Gamma_{Z}^{2}}\Big[ g_{v}^{2}+ g_{a}^{2} \Big]^{2},
\label{27}
\end{equation}
with $\Gamma_{Z}=2.4952\pm 0.0023\,GeV$ is the total width of the $Z$-boson decay. The laser-free partial decay width of the $Z$-boson to pairs of electrons and muons are given by:
\begin{equation}
\Gamma_{e^{+}e^{-}}=\dfrac{g^{4}}{\cos^{4}\theta_{w}}\dfrac{M_{Z}}{48\pi}\Big[ g_{v}^{2}+ g_{a}^{2} \Big],
\label{28}
\end{equation}
\begin{equation}
\Gamma_{\mu^{+}\mu^{-}}=\dfrac{g^{4}}{\cos^{4}\theta_{w}}\dfrac{M_{Z}}{48\pi}\Big[ g_{v}^{2}+ g_{a}^{2} \Big], 
\label{29}
\end{equation}
with {$ g $} is the electroweak coupling constant which is related to the electroweak mixing angle $\theta_{w}$ by $g^{2}=e^{2}/\sin^{2}\theta_{w}=8G_{F}M_{Z}^{2}\cos^{2}_{\theta_{w}}/\sqrt{2}$ where $e=1eV$ and $G_{F}=1.166 3787 \times 10^{-5} GeV^{-2}$ are the electron's charge and the Fermi coupling constant, respectively.
 The TCS can be expressed in terms of partial decay widths $\Gamma_{e^{+}e^{-}}$ and $\Gamma_{\mu^{+}\mu^{-}}$. Thus,
\begin{equation}
\sigma=\frac{12\pi}{M_{Z}^{2}}\dfrac{s}{(s-M_{Z}^{2})^{2}+M_{Z}^{2}\Gamma_{Z}^{2}}\Gamma_{ee}\Gamma_{\mu\mu}.
\label{30}
\end{equation}
Now, we proceed by evaluating the partial width decay of $Z$-boson into pair of electron and positron in the presence of a CP-laser field. These quantity is already evaluated in \cite{7} and it is given by:
\begin{equation}
\Gamma_{e^{+}e^{-}}^{n}=\dfrac{G_{F}M_{Z}}{16\sqrt{2}(2\pi)^{2}}\int\dfrac{|\mathbf{q_{e^{+}}}|^{2}d\Omega}{Q_{e^{-}}Q_{e^{+}}g^{ \prime}(|\mathbf{q_{e^{-}}}|)}\big| \overline{M_{fi}^{n}} \big|^{2},
\label{31}
\end{equation}
with  $\big| \overline{M_{fi}^{n}} \big|^{2} $ is the squared modulus of the scattering amplitude. $g^{ \prime}(|\mathbf{q_{e^{-}}}|)$ is given by:
 \begin{equation}
g^{ \prime}(|\mathbf{q_{e^{-}}}|)=\dfrac{|\mathbf{q_{e^{-}}}|}{\sqrt{|\mathbf{q_{e^{-}}}|^{2}+m_{e}^{*^{2}}}}+\dfrac{|\mathbf{q_{e^{-}}}|-n\omega\cos(\theta)}{\sqrt{(n\omega)^{2}+|\mathbf{q_{e^{-}}}|^{2}-2n\omega|\mathbf{q_{e^{-}}}|\cos(\theta)+m_{e}^{*^{2}}}},
\label{32}
\end{equation}
 $(n)$ is the number of exchanged photons between the colliding system and the laser field. $Q_{{e^{+}}}$ and $Q_{{e^{-}}}$ are the energy of the incident electron and positron, respectively. $\mathbf{q_{e^{-}}}$ is the effective momentum of the electron and $m_{e}^{*}$ is the effective mass of the electron. The integral over $d\Omega=\sin\theta d\theta d\phi$ involved in the evaluation of $ \Gamma_{e^{+}e^{-}}^{n} $ is performed by using a numerical integration.
Finally, the laser-assisted Breit-Wigner's TCS is obtained by replacing the laser-free partial width $ \Gamma_{e^{+}e^{-}} $ with $ \Gamma_{e^{+}e^{-}}^{n} $ in Eq.(\ref{30})
\begin{equation}
\sigma=\frac{12\pi}{M_{Z}^{2}}\dfrac{s}{(s-M_{Z}^{2})^{2}+M_{Z}^{2}\Gamma_{Z}^{2}}\sum_{n=-\infty}^{+\infty}\Gamma_{e^{+}e^{-}}^{n}\Gamma_{\mu\mu}
\label{33}
\end{equation}
\subsection{Results and discussion}
In this part, we present and discuss some collected data for the total and differential cross section of the scattering process $e^{+}e^{-}\rightarrow \mu^{+} \mu^{-} $ in which the colliding particles are embedded in a strong laser field. First, we have checked the validity of our results by comparing the TCS in the presence of a CP-laser field with that obtained for the laser-free process \cite{21}. Next, we have shown the dependence of the TCS on the transferred photons number. Then, we have illustrated the variation of the TCS as a function of the CME for different number of exchanged photons. Later, we have shown how the TCS varies as a function of the laser field's strength for different laser's frequencies. Finally, we have explored the dependence of the differential cross section on the CME for different scattering angles. To calculate out the cross sections numerically, we take the parameters from the PDG \cite{22}: The mixing angle $\sin^{2}(\theta_{w})=0.23126$, the $Z$-boson mass $M_{Z}=91.186\,GeV$ and its total width $\Gamma_{Z}=2.4952\,GeV$.
\begin{figure}[H]
  \centering
      \includegraphics[scale=0.8]{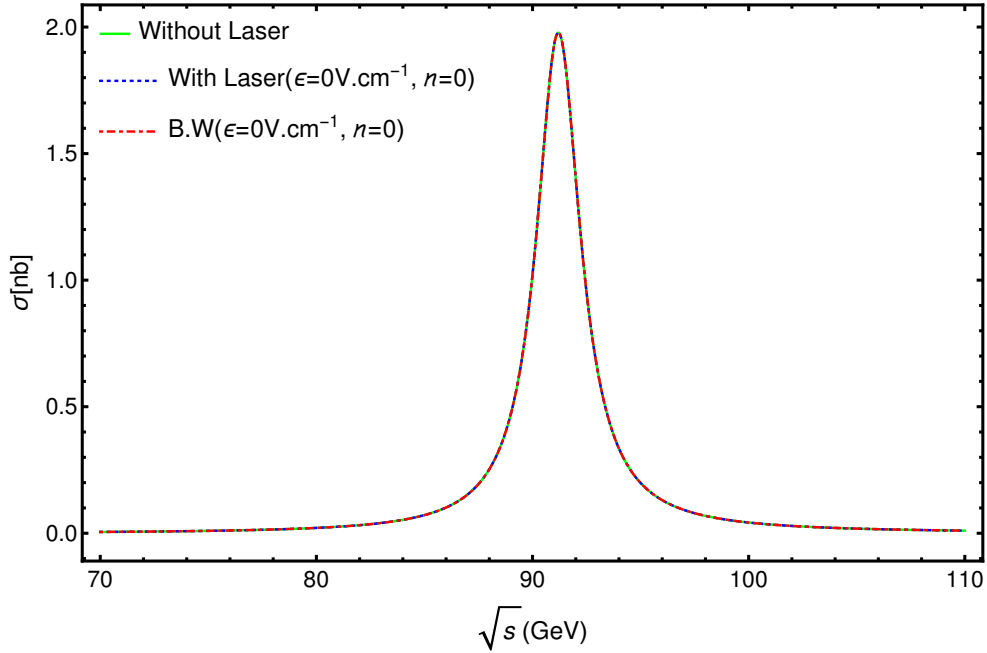}
  \caption{Comparison between the  laser-free TCS \cite{21}, the TCS given by Eq.(\ref{26}) which is obtained by using the scattering matrix approach and the Breit-Wigner's TCS given by Eq.(\ref{33}), in the presence of a CP-laser field (with $\varepsilon_{0}=0\,V.cm^{-1}$ and $n=0$).}
  \label{fig2}
\end{figure}
 Figure \ref{fig2} shows the comparison between the laser-assisted TCS of the $Z$-boson production calculated by using the scattering matrix approach, the Breit-Wigner TCS described by Eq.(\ref{33}) in the presence of the laser field and the corresponding TCS of the laser field-free process. This comparison is aimed to check the validity of our results by taking the laser field's strength $\varepsilon_{0}=0\,V.cm^{-1}$ and the number of exchanged photons $n=0$. We have integrated the Breit-Wigner formula in order to indicate that it can be used not only in laser-free process but also in the presence of a CP-laser field. Apparently, all curves presented in Fig.\ref{fig2} are in excellent agreement for all centre of mass energies. Therefore, the TCS in the presence of the laser field converges to the TCS in the absence of the laser field when the latter's parameters are taken to be zero.
To check precisely the validity of the Breit-Wigner formula in the presence of an external field, we have calculated some numerical values of the TCS expressed by both Eq.(\ref{26}) and Eq.(\ref{33}) where the colliding particles $(e^{+}$ and $e^{-})$ are embedded in a circularly polarized laser beam.
\begin{table}[t]
\centering
\caption{\label{tab1}Comparison between numerical values of the TCS described by Eq.(\ref{26}) and the laser-assisted Breit-Wigner formula given by Eq.(\ref{33}) for different numbers of exchanged photons and laser field's strengths. The CME and the laser's frequency are chosen as $\sqrt{s}=M_{Z}$ and $\hbar\omega=0.117\,eV$, respectively.}
\begin{ruledtabular}
\begin{tabular}{cccc}
Exchanged photons number & $ \varepsilon_{0}[V.cm^{-1}] $ &  TCS[nb] given by Eq.(\ref{26})  &  TCS[nb] given by Eq.(\ref{33}) \\
 \hline
  \multirow{4}{*}{$-50\leq n\leq 50$} 
  & $ 10\rightarrow 10^{4} $ & $ 1.97745 $ & $ 1.97737 $\\
    & $ 10^{5} $ & $ 0.223693 $ & $ 0.223686 $\\
    & $ 10^{6} $ & $ 0.0218601 $ & $ 0.0218594 $\\
    & $ 10^{7} $ & $ 0.00219158 $ & $ 0.00219147 $\\
    & $ 10^{8} $ & $ 0.000218418 $ & $ 0.000218409 $\\ \hline
    \multirow{4}{*}{$-200\leq n\leq 200$}
     &  $ 10\rightarrow 10^{4} $ & $ 1.97745 $ & $ 1.97737 $\\
    & $ 10^{5} $ & $ 0.968742 $ & $ 0.968733 $\\
    & $ 10^{6} $ & $ 0.0875007 $ & $ 0.0875001 $\\
    & $ 10^{7} $ & $ 0.00876803 $ & $ 0.00876795 $\\
    & $ 10^{8} $ & $ 0.000873503 $ & $ 0.000873491 $\\ 
\end{tabular}
\end{ruledtabular}
\end{table}
Table \ref{tab1} shows a comparison between numerical values of the TCS obtained by using the scattering matrix theory and those obtained by using the Breit-Wigner formula in the presence of a CP-laser field. As we are studying the $Z$-boson production, the CME is taken as $\sqrt{s}=M_{Z}$ and the laser's frequency is chosen as $\hbar\omega=0.117\,eV$. According to the Table \ref{tab1}, the two total cross sections, for different electric field amplitudes, are very consistent for that they give the same results, regardless of the number of exchanged photons. This surprising result indicates that the TCS given by the famous Breit-Wigner formula is still valid in the presence of a CP-laser field. In addition, for the number of exchanged photons $-50\leq n\leq 50$ and $-200\leq n\leq 200$, the CP-laser field doesn't affect the TCS unless its electric field amplitude reaches $\varepsilon_{0}=10^{5}\,V.cm^{-1}$. The reason behind this result is that the electron's effective mass is equal to the corresponding electron's free mass $(m_{e}^{*}=m_{e}=0.511\times 10^{-3}\,GeV)$ for the laser field's strengths ($\varepsilon_{0}$) which are included in the interval [$10\,V.cm^{-1}$, $10^{4}\,V.cm^{-1}$]. To be more specific, the term ($e^{2}a^{2}$), that appears in Eq.(\ref{7}), significantly increases the effective mass of the electron and positron inside the laser field, especially when the laser field's strength reaches a threshold  $\varepsilon_{0}=10^{5}\,V.cm^{-1}$. Let's move now to study the dependence of the partial total cross section on the net exchanged photons number $n$.
\begin{figure}[h]
  \centering
      \includegraphics[scale=0.65]{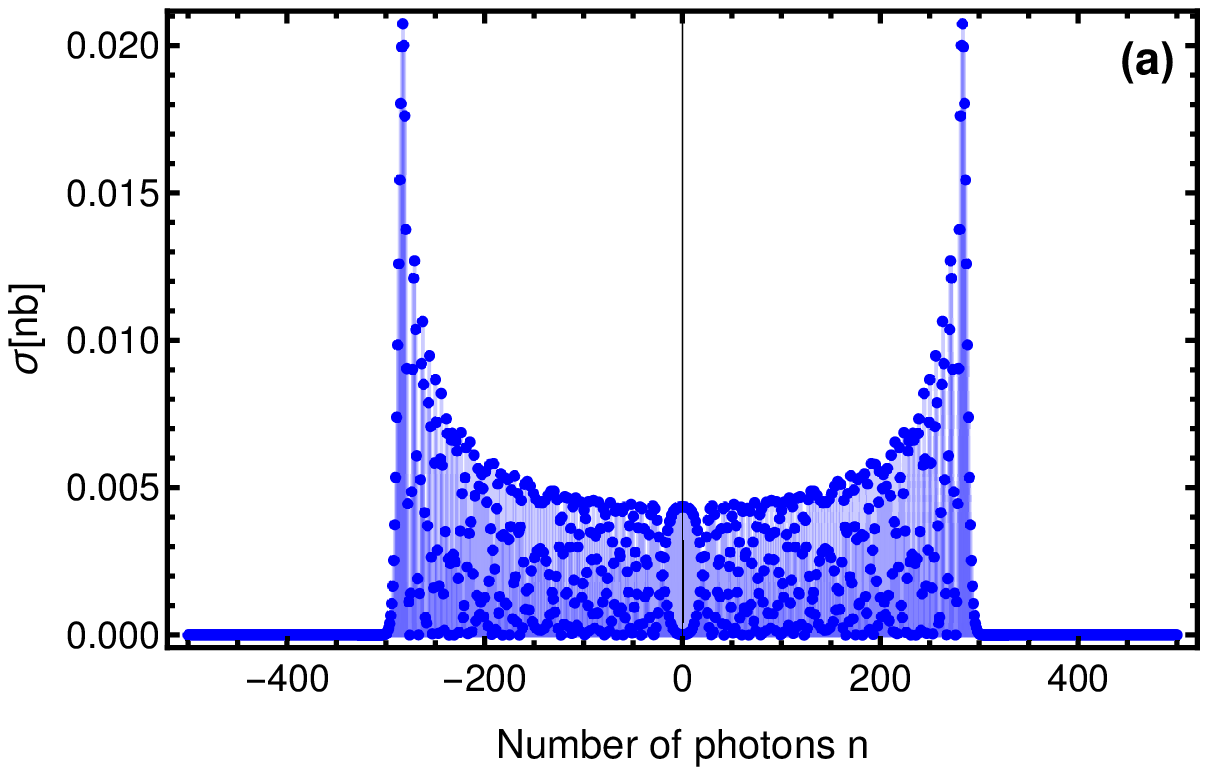}\hspace*{0.4cm}
      \includegraphics[scale=0.65]{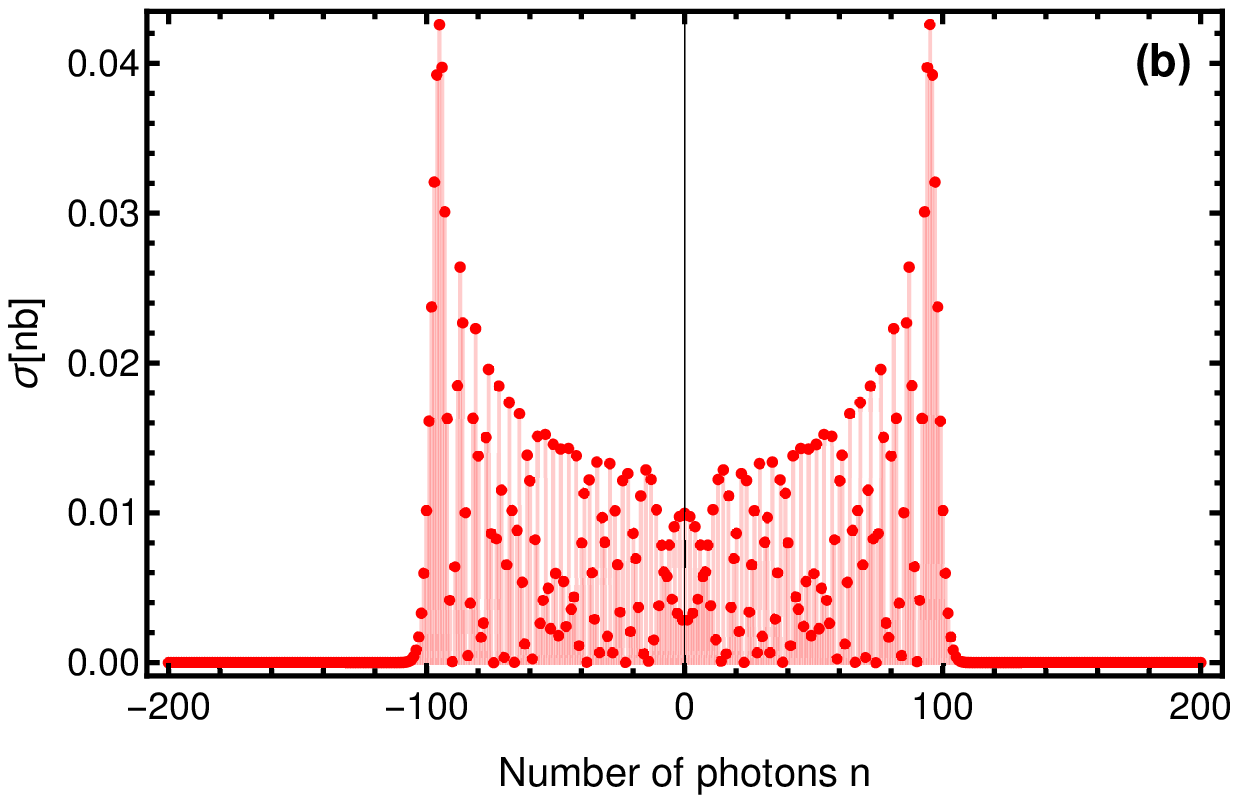}\par\vspace*{0.5cm}
      \includegraphics[scale=0.65]{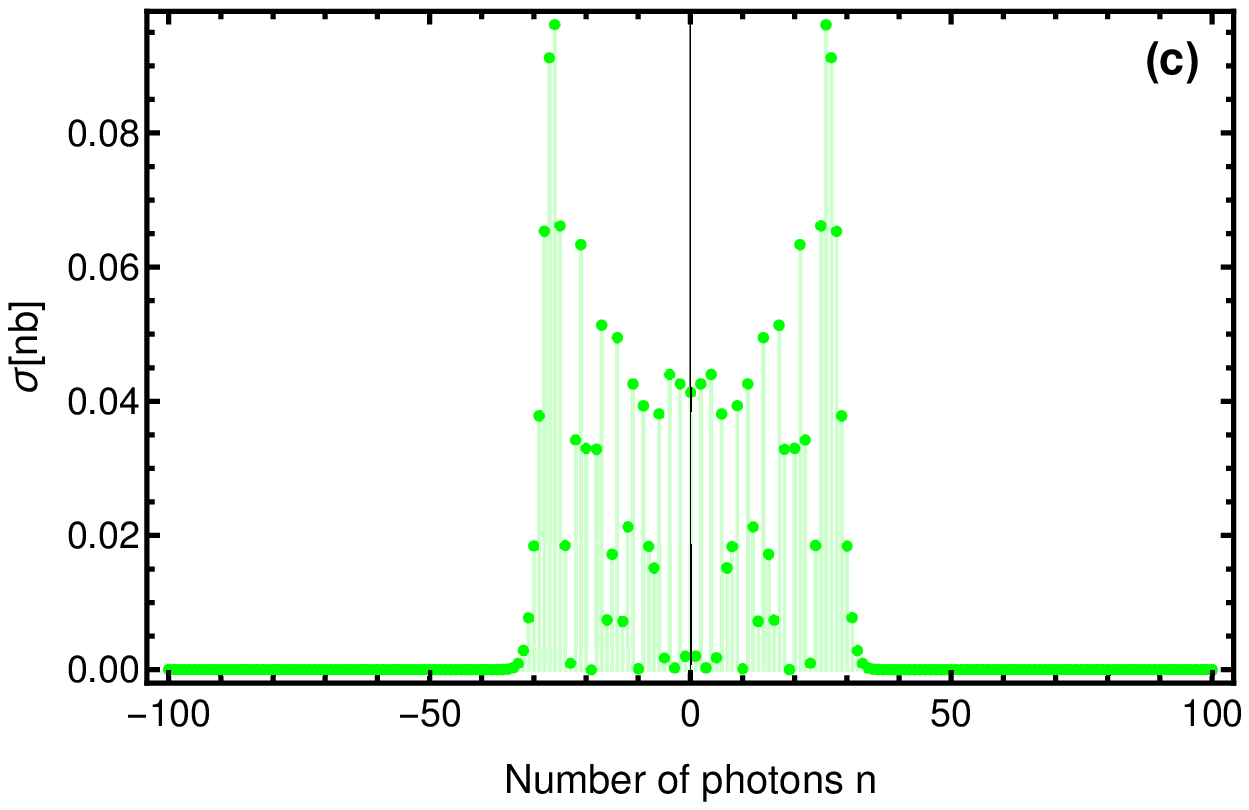}
        \caption{The TCS of the $Z$-boson production via $e^{+}e^{-}\rightarrow \mu^{+} \mu^{-} $ in natural units as a function of the transferred photons' number for the CME $(\sqrt{s}=M_{Z})$.  The laser field's strengths in Figs.\ref{fig3}{\color{blue}(a)}, \ref{fig3}{\color{blue}(b)} and \ref{fig3}{\color{blue}(c)} are successively  $\epsilon_{0}=10^{7}V.cm^{-1}$ , $\epsilon_{0}=10^{7}V.cm^{-1}$ and $\epsilon_{0}=10^{6}V.cm^{-1}$. The laser field frequencies in Figs.\ref{fig3}{\color{blue}(a)}, \ref{fig3}{\color{blue}(b)} and \ref{fig3}{\color{blue}(c)} are $\omega=1.17\, eV$, $\omega=2\, eV$ and $\omega=1.17\, eV$, respectively.}
        \label{fig3}
\end{figure}
Figures \ref{fig3}{\color{blue}(a)}, \ref{fig3}{\color{blue}(b)} and \ref{fig3}{\color{blue}(c)} illustrate the partial TCS of muon pair production, via $Z$-boson exchange, as a function of the net photon number $n$ transferred between the laser field and the colliding system. The laser field's amplitude and frequency in Fig.\ref{fig3}{\color{blue}(a)}, Fig.\ref{fig3}{\color{blue}(b)} and Fig.\ref{fig3}{\color{blue}(c)} are successively taken as ($\varepsilon_{0}=10^{7}V.cm^{-1}$ ; $\hbar\omega=1.17\, eV$), ($\varepsilon_{0}=10^{7}V.cm^{-1}$; $\hbar\omega=2\, eV$) and ($\varepsilon_{0}=10^{6}V.cm^{-1}$; $\hbar\omega=1.17\, eV$). The magnitude of the TCS varies in the range of few orders for different $n$. Furthermore, the contribution of various $n$-photon processes prsesnts cutoffs  at two edges. These cutoffs are symmetric with respect to $n = 0$. In addition, all figures present oscillations which are inherent to the presence of Bessel functions. In general, the physical system has to exchange a cutoff number of photons in order to reach the kroll-Whatson sum-rule \cite{18}, which is described by Eq.(\ref{26}). This cutoff number strongly depends on the laser field's strength and its frequency. According to Fig.\ref{fig3}{\color{blue}(a)}, Fig.\ref{fig3}{\color{blue}(b)} and Fig.\ref{fig3}{\color{blue}(c)}, this cutoff number is approximately $-300\leq n\leq +300$, $-110\leq n\leq +110$ and $-33\leq n\leq +33$, respectively. By comparing Fig.\ref{fig3}{\color{blue}(a)} and Fig.\ref{fig3}{\color{blue}(b)}, we observe that the cutoff number decreases as far as the laser's frequency increases. In contrast, for the same laser's frequency as in Fig.\ref{fig3}{\color{blue}(b)} and Fig.\ref{fig3}{\color{blue}(c)}, the required photons number to reach the sum-rule rises by increasing the laser field's strength. Now, Let's focus our attention on how the TCS varies as a function of the CME for different number of exchanged photons.
\begin{figure}[t]
  \centering
      \includegraphics[scale=0.8]{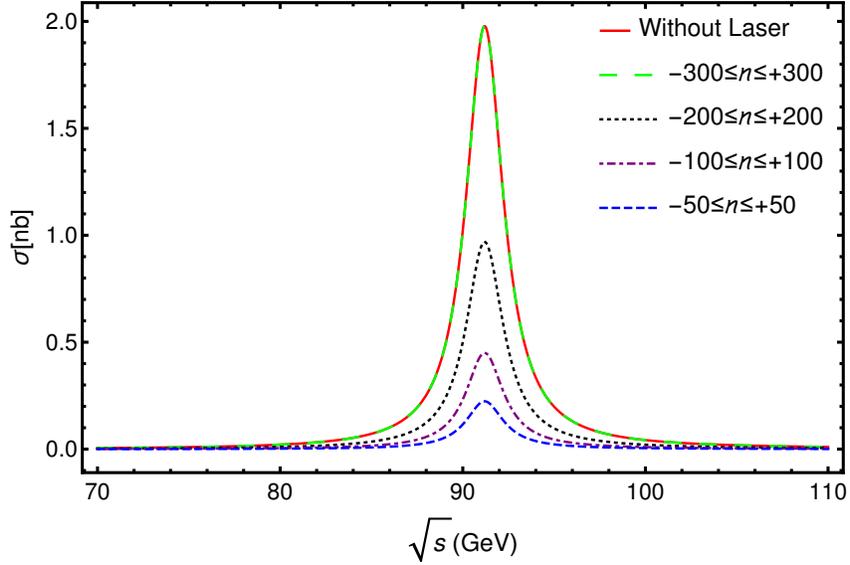}
  \caption{Variation of the TCS of the $Z$-boson production via $e^{+}e^{-}\rightarrow \mu^{+} \mu^{-} $ in the presence of a circularly polarized laser beam as a function of  the CME $(\sqrt{s})$ for different number of exchanged photons. The laser field's strength and its frequency are taken as $\varepsilon_{0}=10^{7}V.cm^{-1}$ and $\hbar\omega=1.17\,eV$, respectively.}
  \label{fig4}
\end{figure}
It is well known that the $Z$-boson production is an important background for many other physics processes and therefore needs to be well understood. Figure \ref{fig4} displays the laser-free TCS given by Eq.(\ref{27}) and the TCS of the process $e^{+}e^{-}\rightarrow \mu^{+}\mu^{-}$, which is given by Eq.(\ref{26}), in the presence of a CP-laser field for different number of exchanged photons. It is obvious that, regardless of the number of exchanged photons, the TCS is negligible for low energies. But it rises when the CME ovecomes a threshold value which is approxiamtely $80\,GeV$. Furthermore, the maximum of the cross section of this resonant process occurs at $\sqrt{s}=M_{Z}$. However, for the centre of mass energies greater than $M_{Z}$ $(\sqrt{s}>M_{Z})$, the cross section falls down abruptly until it becomes zero. In \cite{14}, the authors showed that The CP-laser field reduces the TCS of the QED process $e^{+}e^{-}\rightarrow \mu^{+} \mu^{-}$. In the present work, we have also found that the CP-laser field decreases the TCS of the weak process $e^{+}e^{-}\rightarrow \mu^{+} \mu^{-}$ by several orders of magnitude, moreover, this decreasing behavior depends on the number of transferred photons $n$. For example, for the CME $\sqrt{s}=M_{Z}$, the maximum of the cross section is approximately $0.2nb$, $0.4nb$, $0.9 nb$  for the number of exchanged photons $-50\leq n\leq +50$, $-100\leq n\leq +100$ and $-200\leq n\leq +200$, respectively. Theoretically, the maximum value of the TCS can be obtained by using the laser-assisted Breit-Wigner formula given by Eq.(\ref{33}), which gives the same results as the Eq.(\ref{26}). Thus, we have, for $\sqrt{s}=M_{Z}$:
 \begin{equation}
\sigma_{max}=\dfrac{12\pi}{M_{Z}^{2}}\dfrac{\sum_{n=-\infty}^{+\infty}\Gamma_{e^{+}e^{-}}^{n}\Gamma_{\mu\mu}}{\Gamma_{Z}^{2}}.
\end{equation}
In addition, when the number of exchanged photons reaches the cutoff number $n=\pm 300$ (Fig.\ref{fig3}{\color{blue}(a)}), the TCS converges to the laser-free cross section. So, the Kroll Whatson sum rule, which is expressed by Eq.(\ref{26}), is checked. Therefore, the CP-laser field reduces the $Z$-boson production cross section with a factor depending on the number of exchanged photons $n$, the laser field's strenth $\varepsilon_{0}$ and its frequency $\hbar\omega$. This effect is symmetric below the $Z$-boson pole and it doesn't shift the location of the cross section's peak  as shown in Fig.\ref{fig4}. Regardless of the number of exchanged photons, the TCS falls down to the half of its peak at value $\sqrt{s}=M_{z}\pm (\Gamma_{Z}/2)$. Then, $\Gamma_{Z}$ is not only the total decay rate of the $Z$-boson but also the full-width at half-maximum of the TCS as a function of the CME \cite{21}. Since $M_{Z}$ and $\Gamma_{Z}$ are well known, the measured value of the peak cross section for muon pair production $\sigma_{max}$ can be related to the product of the partial decay widths by: $\sum_{n=-\infty}^{+\infty}\Gamma_{e^{+}e^{-}}^{n}\Gamma_{\mu\mu}=\sigma_{max}M_{Z}^{2}\Gamma_{Z}^{2}/12\pi$. Due to the fact that the CP-laser field reduces the maximum cross section ($\sigma_{max}$), we assume that the partial width of the $Z$-boson decay into electron positron pair production $\sum_{n=-\infty}^{+\infty}\Gamma_{e^{+}e^{-}}^{n}=(M_{Z}^{2}\Gamma_{Z}^{2}/12\pi)\times (\sigma_{max}/\Gamma_{\mu\mu})$ decreases as a function of the number of exchanged photons for a given laser field's strength and frequency. This result confirms the same behavior obtained in\cite{7}, in which the authors show that, at high laser intensity, the CP-laser field decreases the branching ratio of the hadronic and leptonic decay modes of the $Z$-boson while its invisible decay mode increases. After having discussed the variation of the TCS as a function of the CME for different number of transferred photons, we next focus on how this TCS varies at $Z$-boson peak $(\sqrt{s}=M_{Z})$ as a function of the laser field's strength for different laser field's frequencies.
\begin{figure}[t]
  \centering
      \includegraphics[scale=0.65]{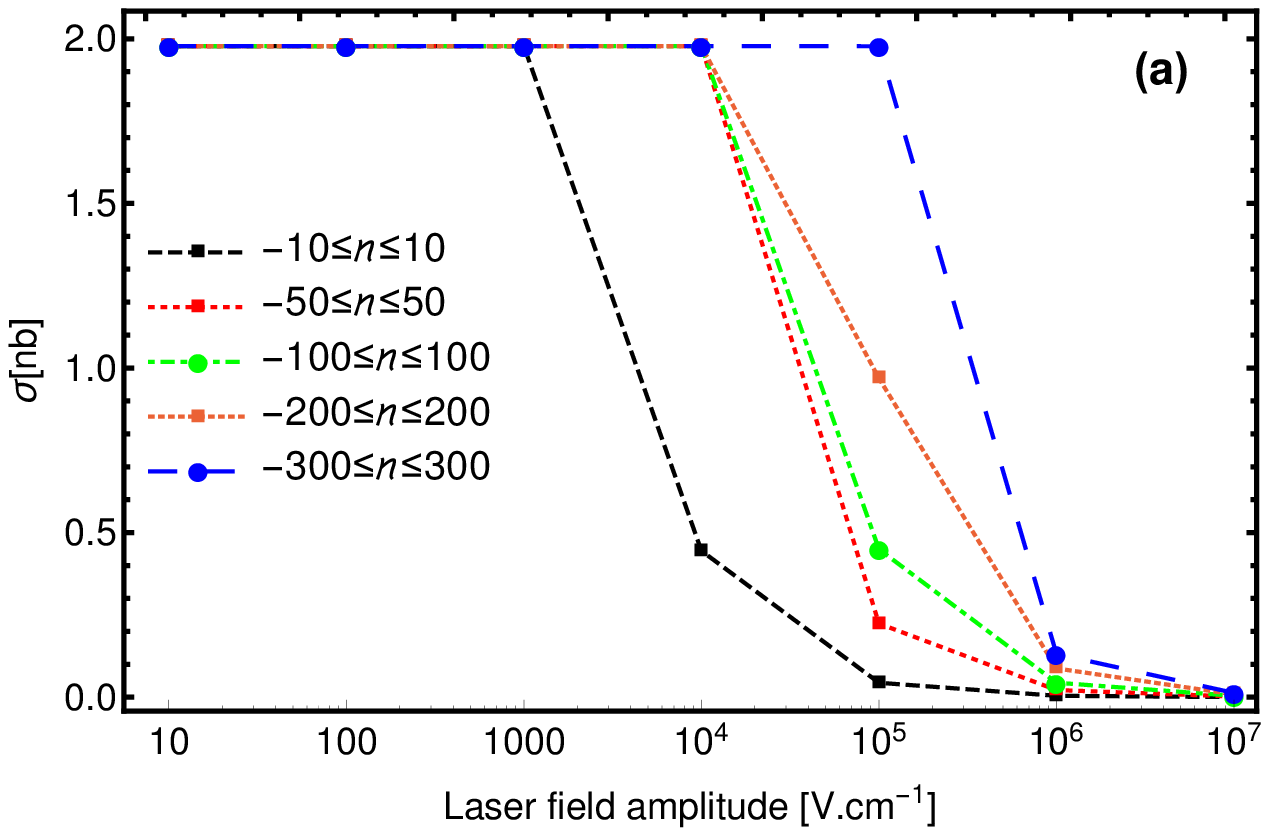}\hspace*{0.4cm}
      \includegraphics[scale=0.65]{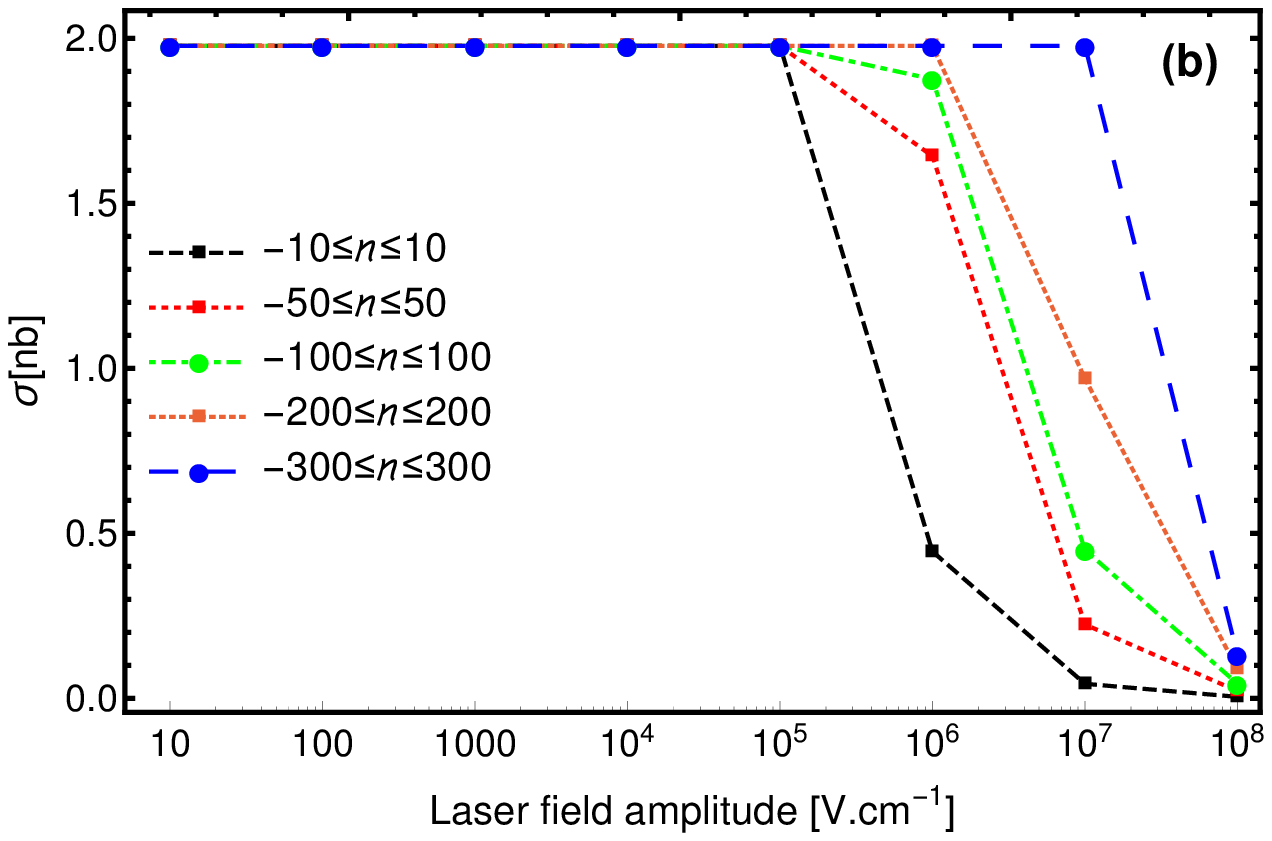}\par\vspace*{0.5cm}
      \includegraphics[scale=0.65]{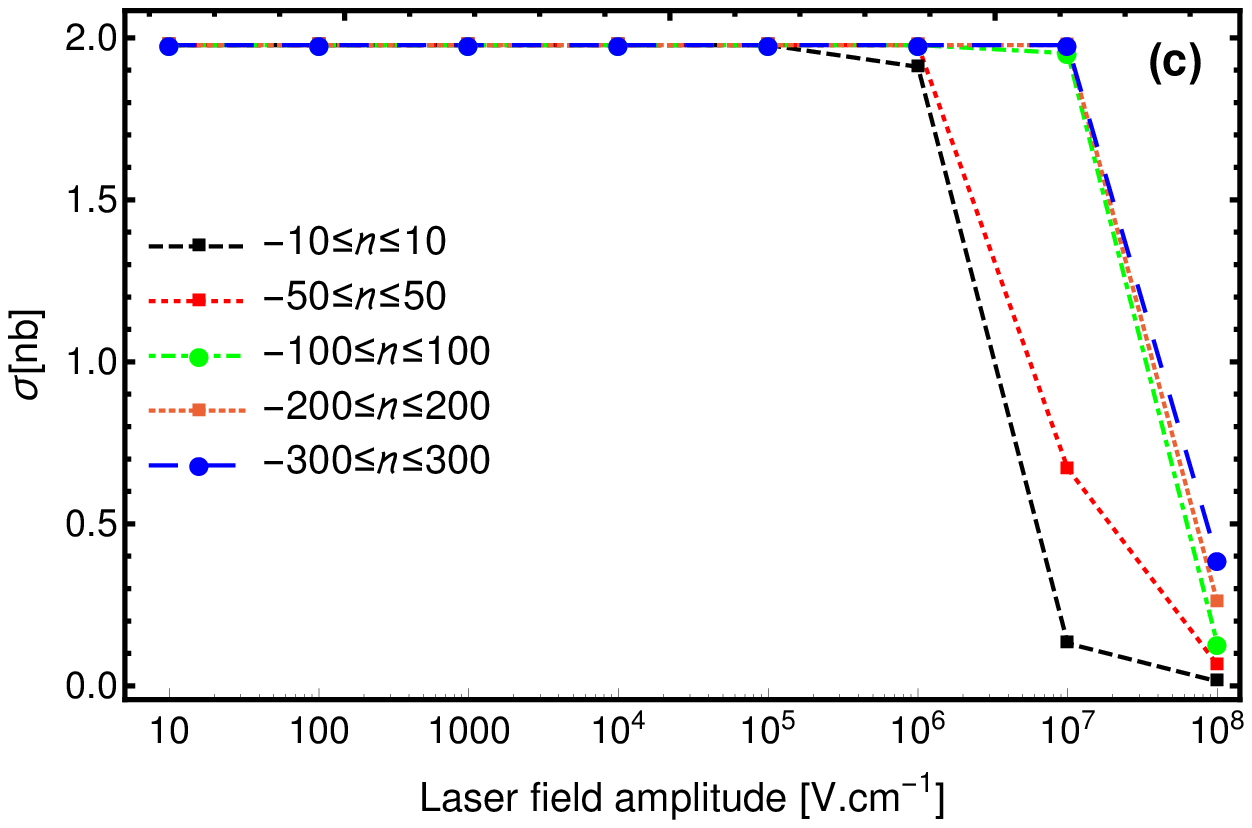}
        \caption{Dependence of the laser-assisted TCS of the $Z$-boson production via $e^{+}e^{-}\rightarrow \mu^{+} \mu^{-} $ on the laser field's strength for different exchanged photons number. The center of mass energy is taken as $\sqrt{s}=M_{Z}$ for all curves. The laser's frequencies in Fig.\ref{fig5}{\color{blue}(a)}, Fig.\ref{fig5}{\color{blue}(b)} and Fig.\ref{fig5}{\color{blue}(c)} are successively chosen as $\,\omega=0.117 eV$, $\omega=1,17 eV$ and $\omega=2 eV$.}
        \label{fig5}
\end{figure}
Figures \ref{fig5}{\color{blue}(a)}, \ref{fig5}{\color{blue}(b)} and \ref{fig5}{\color{blue}(c)} show the laser-assisted TCS versus the laser field's amplitude for the laser's frequencies $0.117\,eV$, $1.17\,eV$ and $2\,eV$, respectively. For small intensities, the TCS doesn't shows any dependence at all on the laser field. The effect of the CP-laser field on the cross section begins at $10^{3}\,V.cm^{-1}$, $10^{5}\,V.cm^{-1}$ and $10^{5}\,V.cm^{-1}$ for the laser's field frequencies $0.117\,eV$, $1.17\,eV$ and $2\,eV$, respectively, for the exchanged photons number $-50\leq n\leq +50$. As we can see from these figures, for each laser's frequency, the threshold value from which the laser field begins to influence the TCS varies according to the number of the transferred photons between the electromagnetic field and the colliding particles. In general, this threshold value rises as far as the laser's frequency or the number of exchanged photons increase. To understand more clearly this dependence of the TCS on the laser field's strength, we have plotted, in one figure, the variation of the TCS as a function of the laser field's strength for different known laser field's frequencies. 
\begin{figure}[h]
  \centering
      \includegraphics[scale=0.8]{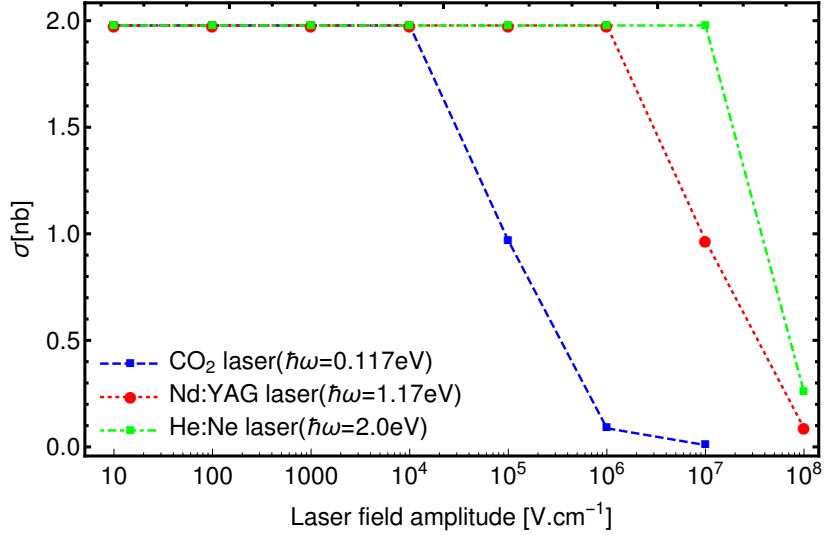}
  \caption{Variation of the TCS of the $Z$-boson production via $e^{+}e^{-}\rightarrow \mu^{+} \mu^{-} $ as a function of the laser field's strength for the CO2 laser, Nd:YAG laser and the He:Ne laser. The CME and the number of exchanged photons are successively taken as $\sqrt{s}=M_{Z}$ and $-200\leq n\leq 200$.} 
  \label{fig6}
\end{figure}
Figure \ref{fig6} displays the dependence of the TCS of the process $e^{+}e^{-}\rightarrow \mu^{+} \mu^{-} $ on the laser field's amplitude $\varepsilon_{0}$ for three known lasers which are the CO2 laser $(\hbar\omega = 0.117 eV)$, the Nd:YAG laser  $(\hbar\omega= 1.17 eV)$ and the $He:Ne$ laser $(\hbar\omega= 2 eV)$. For the number $-200\leq n\leq 200$ of exchanged photons between the laser field and the incident particles, the effect of the CO2 laser, Nd:YAG laser and the He:Ne laser, begins to appear at $\varepsilon_{0}=10^{4}V.cm^{-1}$, $\varepsilon_{0}=10^{6}V.cm^{-1}$ and $\varepsilon_{0}=10^{7}V.cm^{-1}$, respectively. This means that for the same number of exchanged photons, the threshold value of the laser field's amplitude increases with respect to the laser's frequency. In addition, all of them highly reduce the TCS until it becomes zero. This result confirms those obtained in Fig.\ref{fig5}{\color{blue}(a)}, Fig.\ref{fig5}{\color{blue}(b)} and Fig.\ref{fig5}{\color{blue}(c)}.
After understanding the dependence of the total cross section on the laser field's strength, we move now to study how it behaves with respect to the laser field's frequency for a given laser field's strength.
\begin{figure}[t]
  \centering
  \includegraphics[scale=1]{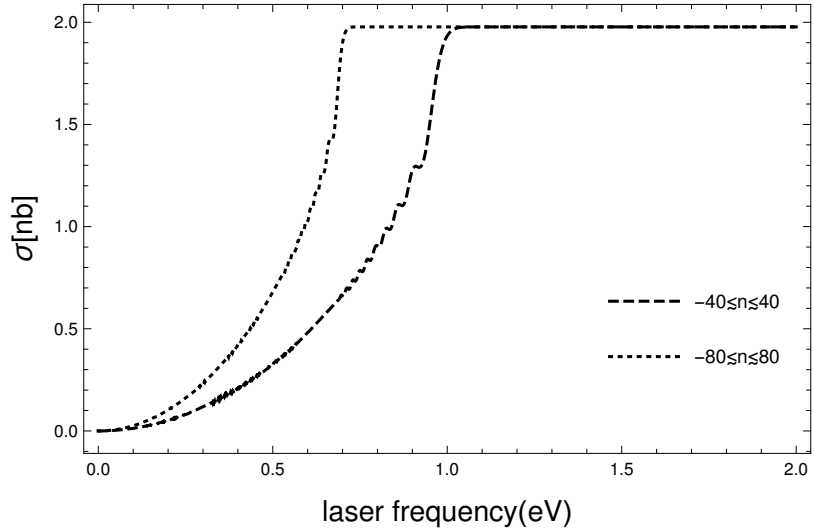}
  \caption{The laser-assisted TCS of the $Z$-boson production via $e^{+}e^{-}\rightarrow \mu^{+} \mu^{-} $  as a function of the laser field's frequency for different number of transferred photons. The CME and the laser field's strength  are successively taken as $\sqrt{s}=M_{Z}$ and $\varepsilon_{0}=\,10^{6}\,V.cm^{-1}$.} 
  \label{fig7}
\end{figure}
Figure \ref{fig7} illustrates the dependence of the TCS on the laser field's frequency $\hbar\omega$ at the $Z$-boson pole for two different numbers of exchanged photons at the interacting particles' vertex. According to Fig.\ref{fig7}, The TCS rises as much as the frequency of the laser field increases. Furthermore, its order of magnitude for each frequency depends on the number of exchanged photons $(n)$. It is obvious that, for each number of exchanged photons and as the laser's frequency reaches a certain value, the laser-assisted TCS becomes comparable to the corresponding laser-free total cross section. These threshold values of the laser's frequencies are $0.7127\,eV$ and $1.006\,eV$ for the transferred photons numbers $-80\leq n\leq 80$ and $-40\leq n\leq 40$, respectively. Moreover, This result is in full agreement with that found in Fig.\ref{fig3}{\color{blue}(c)}.\\
The scattering differential cross-section is a vital operator to understand how the scattering angle affects the physical interaction and the physical properties of the system. In this sense, we have discussed the dependence of the summed differential cross section ($d\sigma/d\Omega$) of the weak process $e^{+}e^{-}\rightarrow \mu^{+} \mu^{-} $ on the CME for different values of the scattering angles ($\theta$ in degrees) in the presence of a circularly polarized laser field.
\begin{figure}[h]
  \centering
  \includegraphics[scale=0.8]{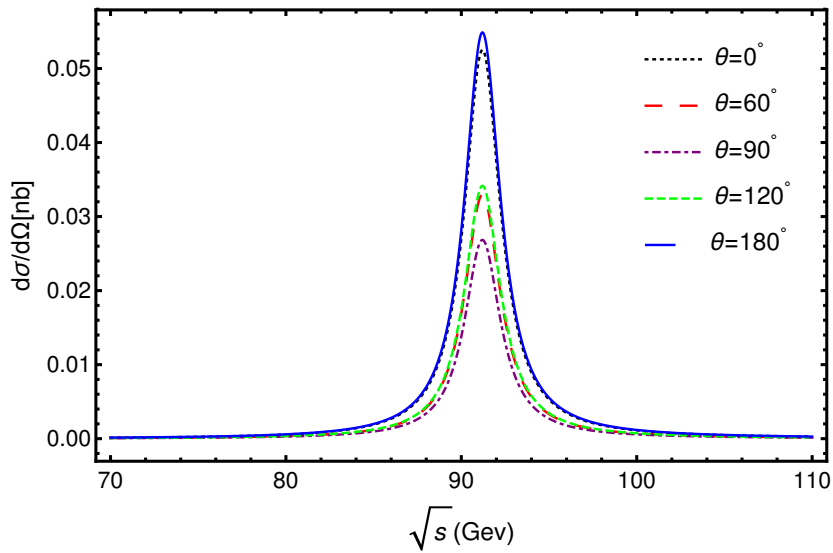}
  \caption{The laser-assisted differential cross section of the $Z$-boson production via $e^{+}e^{-}\rightarrow \mu^{+} \mu^{-} $ process as a function of the CME $(\sqrt{s})$. The number of exchanged photons is $-100\leq n\leq 100$. The laser field's strength and its frequency are successively chosen as $\varepsilon_{0}=\,10^{7}\,V.cm^{-1}$ and $\hbar\omega=1.17\,eV$.} 
  \label{fig8}
\end{figure}
Figure \ref{fig8} illustrates the variation of the laser-assisted differential cross section as a function of the CME for five scattering angles $(\theta=0\degree,\,\theta=60\degree,\,\theta=90\degree,\,\theta=120\degree,\,\theta=180\degree)$. The curves present a peak (resonance) at the CME $\sqrt{s}=M_{Z}$, regardless of the value of the scattering angle $\theta$. To avoid intensive computation, we have chosen the number of exchanged photons as $-100\leq n\leq 100$. The laser's electric field amplitude and its frequency are taken as $\varepsilon_{0}=10^{7}\,V.cm^{-1}$ and $\hbar\omega=1.17\,eV$, respectively. According to Fig.\ref{fig8}, it is clear that the maximum of the differential cross section occurs for low scattering angles $(\theta \rightarrow 0\degree)$ and high scattering angles $(\theta \rightarrow 180\degree)$. However, for the scattering angles near $\theta\approx 90\degree $, the differential cross section reaches its minimum.
\section{ $Z$-boson production via $e^{+}e^{-}\rightarrow \mu^{+} \mu^{-}$ in the presence of a CP-laser field} \label{section 3}
\subsection{Theoretical framework } \label{subsection 3}
This paper's part is devoted to the analytical calculation of the total cross section of the process ${e}^{+} {e}^{-} \rightarrow {\mu}^{+} {\mu}^{-}$ in the presence of a CP-laser field. For this reason, we have to proceed in the same manner as in Sec.\,\ref{section 2} and taking into consideration the fact that both incident and scattered charged particles are dressed with the electromagnetic field which is described in Subsec.\,\ref{subsection 1}.
\subsubsection{Evaluation of the TCS by using S-matrix scattering element}
In the S-matrix element given by the Eq.(\ref{2}), we replace the free wave functions of the muon and the antimuon given by Eq.(\ref{6}) by Dirac-Volkov states given by:
\begin{equation}
\begin{cases}
\psi_{p_{3},s_{1}}(y)= \Big[1-\dfrac{e \slashed k \slashed A}{2(k.p_{3})}\Big] \frac{u(p_{3},s_{3})}{\sqrt{2Q_{3}V}} e^{iS(q_{3},s_{3})} \\
\psi_{p_{4},s_{4}}(y)= \Big[1+\dfrac{e \slashed k \slashed A}{2(k.p_{4})}\Big] \frac{v(p_{4},s_{4})}{\sqrt{2Q_{4}V}} e^{iS(q_{4},s_{4})}
\end{cases},
\label{35}
\end{equation}
with
\begin{equation}
\begin{cases}
S(q_{3},s_{3})=-q_{3}x +\frac{e(a_{1}.p_{3})}{k.p_{3}}\sin\phi - \frac{e(a_{2}.p_{3})}{k.p_{3}}\cos\phi\\
S(q_{4},s_{4})=+q_{4}x +\frac{e(a_{1}.p_{4})}{k.p_{4}}\sin\phi - \frac{e(a_{2}.p_{4})}{k.p_{4}}\cos\phi
\end{cases} .
\label{36}
\end{equation}
$Q_{3}$ and $Q_{4}$ are the energies acquired by the muon and the antimuon in the presence of the laser field.
 After necessary replacements, we find that:
\begin{multline}
S_{fi}= -i\dfrac{g^{2}}{16 \cos^{2}\theta_{w}}   \dfrac{1}{\sqrt{16Q_{1}Q_{2}Q_{3}Q_{4}V^{4}}}   \sum_{N=-\infty}^{+\infty} \sum_{n=-\infty}^{+\infty} \Bigg[  \dfrac{1}{(q_{1}+q_{2}+nk)^{2}-M_{Z}^{2}}  \Bigg] M_{fi}^{(n,N)}\,\\(2\pi)^{4}\delta^{4}(q_{3}+q_{4}-q_{1}-q_{2}-(n+N)k),
\label{37}
\end{multline}
with
$N$ and $n$ are the number of emitted $(n,N<0)$ or absorbed $(n,N> 0)$ photons at the electron and muon vertex, respectively. $M_{fi}^{(n,N)}$ is the scattering amplitude, which is defined by:
 \begin{equation}
 M_{fi}^{(n,N)}=\Big[  \overline{v}(p_{2},s_{2})\Gamma^{n}_{\mu} u(p_{1},s_{1})   \Big]     \Big[   \overline{u}(p_{3},s_{3})\Gamma^{N}_{\mu} v(p_{4},s_{4}) \Big],
 \label{38}
 \end{equation}
where $\Gamma^{n}_{\mu}$ is defined by Eq.(\ref{10}) and $\Gamma^{N}_{\mu}$ is given by:
\begin{equation}
\Gamma^{N}_{\mu}=R^{0}_{\mu}\,G_{0N}(z^{\prime})+R^{1}_{\mu}\,G_{1N}(z^{\prime})+R^{2}_{\mu}\,G_{2N}(z^{\prime}).
\label{39}
\end{equation}
The coefficients that appears in $\Gamma^{N}_{\mu}$ are explicitly expressed, in terms of Bessel functions, as follows:
\begin{equation}
 \left.
  \begin{cases}
      G_{0N}(z^{\prime} ) \\
      G_{1N}(z^{\prime} ) \\
      G_{2N}(z^{\prime} )
  \end{cases}
  \right\} = \left.
  \begin{cases}
     J_{N}(z^{\prime} )e^{iN\phi _{0}^{\prime}}\\
    \frac{1}{2}\big(J_{N+1}(z^{\prime} )e^{i(N+1)\phi _{0}^{\prime}}+J_{N-1}(z^{\prime} )e^{i(N-1)\phi _{0}^{\prime}}\big) \\
     \frac{1}{2i}\big(J_{N+1}(z^{\prime} )e^{i(N+1)\phi _{0}^{\prime}}-J_{N-1}(z^{\prime} )e^{i(N-1)\phi _{0}^{\prime}}\big)
  \end{cases}
  \right\} ,
  \label{40}
\end{equation}
and
\begin{equation}
\begin{cases}R^{0}_{\mu}=\gamma_{\mu}(g_{v}-g_{a}\gamma^{5})+2c_{p_{3}}c_{p_{4}}a^{2}k_{\mu}\slashed k(g_{v}-g_{a}\gamma^{5})   &\\R^{1}_{\mu}=c_{p_{3}}\slashed a_{1}\slashed k\gamma_{\mu}(g_{v}-g_{a}\gamma^{5})-c_{p_{4}}\gamma_{\mu}(g_{v}-g_{a}\gamma^{5})\slashed k  \slashed a_{1}   &\\R^{2}_{\mu}=c_{p_{3}}\slashed a_{2}\slashed k\gamma_{\mu}(g_{v}-g_{a}\gamma^{5})-c_{p_{4}}\gamma_{\mu}(g_{v}-g_{a}\gamma^{5})\slashed k  \slashed a_{2}\end{cases}.
\label{41}
\end{equation}
The argument of the Bessel functions is expressed by: 
$z^{\prime}=\sqrt{\alpha_{3}^{2}+\alpha_{4}^{2}}$ and\, $\phi_{0}^{\prime}= \arctan(\alpha_{4}/\alpha_{3})$, where
\begin{center}
$\alpha_{3}=\dfrac{e(a_{1}.p_{3})}{(k.p_{3})}-\dfrac{e(a_{1}.p_{4})}{(k.p_{4})}$  \hspace*{1cm};\hspace*{1cm} $\alpha_{4}=\dfrac{e(a_{2}.p_{3})}{(k.p_{3})}-\dfrac{e(a_{2}.p_{4})}{(k.p_{4})}$.\\
\end{center}
 By using the standard techniques of the differential cross section's calculation, we have found:
 \small
\begin{multline}
\dfrac{d\sigma_{n,N}}{d\Omega}=  \dfrac{g^{4}}{256\times 4\cos^{4}\theta_{w}} \dfrac{1}{\sqrt{(q_{1}q_{2})^{2}-m_{e}^{*^{4}}}}  \sum_{N=-\infty}^{+\infty}\sum_{n=-\infty}^{+\infty} \Big[  \dfrac{1}{((q_{1}+q_{2}+nk)^{2}-M_{Z}^{2})^{2}+M_{Z}^{2}\Gamma_{Z}^{2} } \Big]  |\overline{M_{fi}^{(n,N)}}|^{2}  \\ \dfrac{2|\textbf{q}_{3}|^{2}}{(2\pi)^{2}Q_{3}} \dfrac{1}{\big|g^{'}(|\textbf{q}_{3}|)\big|}, 
\label{42} 
\end{multline}
\normalsize
where 
\begin{equation}
\big|g^{'}(|\textbf{q}_{3}|)\big|=-2\Bigg[\Big[ \sqrt{s}+(n+N)\omega+\frac{e^{2}a^{2}}{2}\Big(\dfrac{4}{\sqrt{s}}\Big)\Big]\dfrac{|\textbf q_{3}|}{\sqrt{|\textbf q_{3}|^{2}+m_{\mu}^{*^{2}}}}\Bigg],
\label{43}
\end{equation}
and 
\begin{equation}
|\overline{M_{fi}^{(n,N)}}|^{2}=\frac{1}{4}Tr\big[ (\slashed p_{1}-m_{e})\Gamma^{n}_{\mu}(\slashed p_{2}+m_{e})\overline{\Gamma^{n}_{\nu}}\, \big]Tr\big[ (\slashed p_{3}-m_{\mu})\Gamma^{N}_{\mu}(\slashed p_{4}+m_{\mu})\overline{\Gamma^{N}_{\nu}}\, \big],
\label{44}
\end{equation}
with $\overline{\Gamma^{n}_{\nu}}$ is given by Eq.(\ref{15}) and  $\overline{\Gamma^{N}_{\nu}}$ is explicitly expressed by:
\begin{equation}
\begin{cases}\overline{R^{0}_{\nu}}=\gamma_{\nu}(g_{v}-g_{a}\gamma^{5})+2c_{p_{3}}c_{p_{4}}a^{2}k_{\nu}\slashed k(g_{v}-g_{a}\gamma^{5})   &\\\overline{R^{1}_{\nu}}=c_{p_{3}}\gamma_{\nu}(g_{v}-g_{a}\gamma^{5})\slashed k\slashed a_{1}-c_{p_{4}}\slashed a_{1}\slashed k \gamma_{\nu}(g_{v}-g_{a}\gamma^{5})    &\\\overline{R^{2}_{\nu}}=c_{p_{3}}\gamma_{\nu}(g_{v}-g_{a}\gamma^{5})\slashed k\slashed a_{2}-c_{p_{4}}\slashed a_{2}\slashed k\gamma_{\nu}(g_{v}-g_{a}\gamma^{5})  \end{cases}.
\label{45}
\end{equation}
The trace calculation is performed by using FEYNCALC program. The quantity $|\overline{M_{fi}^{(n,N)}}|^{2}$ can be expressed in terms of Bessel functions as follows:
\small
\begin{eqnarray}
|\overline{M_{fi}^{(n,N)}}|^{2}&=&\nonumber\dfrac{1}{4}[A_{1}\,J_{n}(z)^{2}J_{N}(z^{\prime})^{2}+A_{2}\,J_{n}(z)^{2}J_{N+1}(z^{\prime})^{2}+A_{3}\,J_{n-1}(z)^{2}J_{N}(z^{\prime})^{2}+A_{4}\,J_{n+1}(z)^{2}J_{N-1}(z^{\prime})^{2}\\&+&\nonumber A_{5}\,J_{n-1}(z)^{2}J_{N-1}(z^{\prime})^{2}+A_{6}\,J_{n+1}(z)^{2}J_{N+1}(z^{\prime})^{2}+B_{1}\,J_{n}(z)J_{n-1}(z)J_{N}(z^{\prime})^{2}\\&+&\nonumber B_{2}\,J_{n}^{2}(z)J_{N}(z^{\prime})J_{N-1}(z^{\prime})+B_{3}\,J_{n}(z)J_{n+1}(z)J_{N}(z^{\prime})^{2}+B_{4}\,J_{n-1}(z)J_{n+1}(z)J_{N}(z^{\prime})^{2}\\&+&\nonumber B_{5}\,J_{n}(z)J_{n+1}(z)J_{N-1}(z^{\prime})^{2}+B_{6}\,J_{n}^{2}(z)J_{N}(z)J_{N+1}(z^{\prime})+B_{7}\,J_{n}(z)J_{n-1}(z)J_{N+1}(z^{\prime})^{2}\\&+&\nonumber B_{8}\,J_{n-1}^{2}(z)J_{N}(z)J_{N-1}(z^{\prime})+B_{9}\,J_{n}(z)J_{n-1}(z)J_{N-1}(z^{\prime})^{2}+B_{10}\,J_{n+1}^{2}(z)J_{N}(z^{\prime})J_{N+1}(z^{\prime})\\&+&\nonumber B_{11}\,J_{n}(z)J_{n+1}(z)J_{N+1}(z^{\prime})^{2}+C_{1}\,J_{n}(z)J_{n-1}(z)J_{N}(z^{\prime})J_{N-1}(z^{\prime})\\&+&\nonumber C_{2}\,J_{n}(z)J_{n+1}(z)J_{N}(z^{\prime})J_{N-1}(z^{\prime})+C_{3}\,J_{n}(z)J_{n-1}(z)J_{N-1}(z^{\prime})J_{N+1}(z^{\prime})\\&+&\nonumber C_{4}\,J_{n}(z)J_{n+1}(z)J_{N}(z^{\prime})J_{N+1}(z^{\prime})+C_{5}\,J_{n+1}(z)J_{n-1}(z)J_{N}(z^{\prime})J_{N+1}(z^{\prime})\\&+& C_{6}\,J_{n-1}(z)J_{n+1}(z)J_{N+1}(z^{\prime})J_{N-1}(z^{\prime}) ], 
\label{46}
\end{eqnarray}
\normalsize
where the coefficients $ A_{i}\,(i=1,2,3,4,5,6) $, $ B_{i}\,(i=1,2,3,4,5,6,7,8,9,10,11) $ and  $ C_{i}\,(i=1,2,3,4,5,6) $ are evaluated by using some computational tools. We give, for example, the expression of the first coefficient multiplied by {$ J_{n}(z)^{2}J_{N}(z^{\prime})^{2} $}.
\small
\begin{eqnarray}
A_{1}&=&\nonumber\dfrac{16}{((k.p_{1}) (k.p_{2}) (k.p_{3}) (k.p_{4}))} (a^4 e^4 (g_{a}^{2} + g_{v}^{2}) (g_{a}^{2}+g_{v}^{2}) ((k.p_{1}) (k.p_{2}) + (k.p_{3}) (k.p_{4}))^2\\ &+&\nonumber 
   2 (k.p_{1}) (k.p_{2}) (k.p_{3}) (k.p_{4}) (4 g_{a}^{2} g_{v}^{2} (-(p_{1}.p_{3}) (p_{2}.p_{4}) + (p_{1}.p_{4}) (p_{3}.p_{2})) + 
      g_{v}^{2} (g_{a}^{2} \\&\times &\nonumber(-m_{\mu}^{2} (2 m_{e}^{2} + (p_{1}.p_{2})) + (p_{1}.p_{3}) (p_{2}.p_{4}) + (p_{1}.p_{4}) (p_{3}.p_{2}) + 
            m_{e}^{2} (p_{3}.p_{4})) + 
         g_{v}^{2} (m_{\mu}^{2}\\&\times &\nonumber (2 m_{e}^{2} + (p_{1}.p_{2})) + (p_{1}.p_{3}) (p_{2}.p_{4}) + (p_{1}.p_{4}) (p_{3}.p_{2}) + 
            m_{e}^{2} (p_{3}.p_{4}))) + 
      g_{a}^{2} (g_{a}^{2} (-m_{\mu}^{2}\\&\times &\nonumber (p_{1}.p_{2}) + (p_{1}.p_{3}) (p_{2}.p_{4}) + (p_{1}.p_{4}) (p_{3}.p_{2}) + 
            m_{e}^{2} (2 m_{\mu}^{2} - (p_{3}.p_{4}))) + 
         g_{v}^{2} (m_{\mu}^{2} (p_{1}.p_{2}) \\&+ &\nonumber (p_{1}.p_{3}) (p_{2}.p_{4}) + (p_{1}.p_{4}) (p_{3}.p_{2}) - 
            m_{e}^{2} (2 m_{\mu}^{2} + (p_{3}.p_{4}))))) + 
   a^2 e^2 ((k.p_{1}) (k.p_{2}) + 
      (k.p_{3})\\&\times & (k.p_{4})) (4 g_{a}^{2} g_{v}^{2} ((k.p_{2}) (k.p_{4}) (p_{1}.p_{3}) - (k.p_{2}) (k.p_{3}) (p_{1}.p_{4}) + 
         (k.p_{1}) (k.p_{3}) (p_{2}.p_{4})\\&- &\nonumber (k.p_{1}) (k.p_{4}) (p_{3}.p_{2})) + 
      g_{a}^{2} (g_{a}^{2} ((k.p_{2}) (k.p_{4}) (p_{1}.p_{3}) + 
            (k.p_{3}) (2 (k.p_{4}) (m_{e}^{2} - (p_{1}.p_{2})) \\&+ &\nonumber (k.p_{2}) (p_{1}.p_{4}) + (k.p_{1}) (p_{2}.p_{4})) + 
            (k.p_{1}) (2 (k.p_{2}) m_{\mu}^{2} + (k.p_{4}) (p_{3}.p_{2}) - 2 (k.p_{2}) (p_{3}.p_{4}))) \\&+ &\nonumber 
         g_{v}^{2} ((k.p_{2}) (k.p_{4}) (p_{1}.p_{3}) + 
            (k.p_{3}) (2 (k.p_{4}) (m_{e}^{2} - (p_{1}.p_{2})) + (k.p_{2}) (p_{1}.p_{4}) + (k.p_{1}) (p_{2}.p_{4})) \\&+ &\nonumber 
            (k.p_{1}) ((k.p_{4}) (p_{3}.p_{2}) - 2 (k.p_{2}) (m_{\mu}^{2} + (p_{3}.p_{4}))))) + 
      g_{v}^{2} (g_{a}^{2} ((k.p_{2}) (k.p_{4}) (p_{1}.p_{3}) + 
            (k.p_{3}) \\&\times &\nonumber(-2 (k.p_{4}) (m_{e}^{2} + (p_{1}.p_{2})) + (k.p_{2}) (p_{1}.p_{4}) + (k.p_{1}) (p_{2}.p_{4})) + 
            (k.p_{1}) (2 (k.p_{2}) m_{\mu}^{2} + (k.p_{4})\\&\times &\nonumber (p_{3}.p_{2}) - 2 (k.p_{2}) (p_{3}.p_{4}))) + 
         g_{v}^{2} ((k.p_{2}) (k.p_{4}) (p_{1}.p_{3}) + 
            (k.p_{3}) (-2 (k.p_{4}) (m_{e}^{2} + (p_{1}.p_{2})) \\&+ &\nonumber (k.p_{2}) (p_{1}.p_{4}) + (k.p_{1}) (p_{2}.p_{4})) + 
            (k.p_{1}) ((k.p_{4}) (p_{3}.p_{2}) - 2 (k.p_{2}) (m_{\mu}^{2} + (p_{3}.p_{4}))))))).
            \label{47}
\end{eqnarray}
\normalsize
The total cross section is obtained by summing over the cutoff number of exchanged photons  and numerically integrating  over the solid angle $d\Omega=\sin\theta d\theta d\phi$. Thus, we get the Kroll Whatson formula \cite{18}
\begin{equation}
\sum_{n=-\infty}^{+\infty}\sum_{N=-\infty}^{+\infty}\sigma_{n,N}= \sigma.
\label{48}
\end{equation}
\subsubsection{Breit-Wigner approach for  $Z$-boson production in the presence of the CP-laser field } 
In this subsection, we have calculated the Breit-Wigner TCS by takaing into account the fact that both incident and scattered particles are embedded in a CP-laser field. For this reason, we have replaced the partial width $\Gamma_{\mu\mu}$ in Eq.(\ref{33}) by that obtained in the presence of a CP-laser field \cite{7}, which is given by:
\begin{equation}
\Gamma_{\mu^{+}\mu^{-}}^{N}=\dfrac{G_{F}M_{Z}}{16\sqrt{2}(2\pi)^{2}}\int\dfrac{|\mathbf{q_{\mu^{+}}}|^{2}d\omega}{Q_{\mu^{-}}Q_{\mu^{+}}g^{ \prime}(|\mathbf{q_{\mu^{-}}}|)}\big| \overline{M_{fi}^{N}} \big|^{2},
\label{49}
\end{equation}
where 
 \begin{equation}
g^{ \prime}(|\mathbf{q_{\mu^{-}}}|)=\dfrac{|\mathbf{q_{\mu^{-}}}|}{\sqrt{|\mathbf{q_{\mu^{-}}}|^{2}+m_{\mu^{-}}^{*^{2}}}}+\dfrac{|\mathbf{q_{\mu^{-}}}|-N\omega\cos(\theta)}{\sqrt{(N\omega)^{2}+|\mathbf{q_{\mu^{-}}}|^{2}-2N\omega|\mathbf{q_{\mu^{-}}}|\cos(\theta)+m_{\mu^{-}}^{*^{2}}}}.
\label{50}
\end{equation}
Finally, the Breit-Wigner TCS becomes as follows:
\begin{equation}
\sigma=\frac{12\pi}{M_{Z}^{2}}\dfrac{s}{(s-M_{Z}^{2})^{2}+M_{Z}^{2}\Gamma_{Z}^{2}}\sum_{n=-\infty}^{+\infty}\sum_{N=-\infty}^{+\infty}\Gamma_{e^{+}e^{-}}^{n}\Gamma_{\mu^{+}\mu^{-}}^{N}.
\label{51}
\end{equation}
\subsection{Results and discussion}\label{subsection 4}
In this subsection, we present and discuss the collected data for the present scattering process by taking into account that both incident and scattered particles are embedded in a circularly polarized laser field.
\begingroup
\begin{table}[H]
 \centering
\caption{\label{tab2}Comparison between numerical values of the TCS given by Eq.(\ref{26}), the laser-assisted Breit-Wigner formula (Eq.\ref{51}) and the TCS given by Eq.(\ref{48}), for different number of exchanged photons and for different laser field's amplitudes. The CME and the laser's frequency are successively chosen as $\sqrt{s}=M_{Z}$ and $\hbar\omega=0.117\,eV$.}
\small
\begin{ruledtabular}
\begin{tabular}{ccccc}
 Exchanged photons numbers & $ \varepsilon_{0}[V.cm^{-1}] $ & TCS[nb] (Eq.\ref{26}) & TCS[nb] (Eq.\ref{48})& TCS[nb] (Eq.\ref{51})\\
 \hline
 \multirow{8}{*}{$-50\leq n, N\leq 50$} 
   & $ 10^{5} $ & $ 0.223693 $ &  $ 0.223693 $& $ 0.223686 $\\
    & $ 10^{6} $ & $ 0.0218601 $ & $ 0.0218601 $ & $ 0.0218594 $\\
    & $ 10^{7} $ & $ 0.00219158 $& $ 0.00219158 $ & $ 0.00219147 $\\
    & $ 10^{8} $ & $ 0.000218418 $& $ 0.000218418 $ & $ 0.000218409 $\\
    & $ 10^{9} $ & $ 2.19751\times 10^{-5} $& $ 1.52064\times 10^{-10} $ & $ 1.52057\times 10^{-10} $\\
    & $ 10^{10} $& $ 2.1412\times 10^{-6} $ & $ 1.51432\times 10^{-12} $ & $ 1.51426\times 10^{-12} $\\
    & $ 10^{11} $ & $ 2.22468\times 10^{-7} $& $ 1.51432\times 10^{-14} $ & $1.51424\times 10^{-14} $\\
    & $ 10^{12} $ & $ 2.22688\times 10^{-8} $& $ 1.51591\times 10^{-16} $ & $ 1.51583\times 10^{-16} $\\
    & $ 10^{13} $ & $ 2.2317\times 10^{-9} $& $ 1.51897\times 10^{-18} $ & $ 1.51888\times 10^{-18} $\\
    & $ 10^{14} $& $ 3.03546\times 10^{-10}  $ & $ 1.94848\times 10^{-20}  $ & $ 1.94839\times 10^{-20} $\\ \hline
    \multirow{8}{*}{$-200\leq n, N\leq 200$}
    & $ 10^{5} $ & $ 0.968742 $ & $ 0.968742 $ & $ 0.968733 $\\
    & $ 10^{6} $& $ 0.0875007 $ & $ 0.0875007 $ & $ 0.0875001 $\\
    & $ 10^{7} $& $ 0.00876803 $ & $ 0.00876803 $ & $ 0.00876795 $\\
    & $ 10^{8} $& $ 0.000873503 $ & $ 0.000873503 $ & $ 0.000873491 $\\
    & $ 10^{9} $ & $ 8.74764\times 10^{-5} $& $ 2.3928\times 10^{-9} $ & $ 2.3922\times 10^{-9} $\\
    & $ 10^{10} $ & $ 8.76396\times 10^{-6} $& $ 2.3903\times 10^{-11} $ & $ 2.3901\times 10^{-11} $\\
    & $ 10^{11} $ & $ 8.77454\times 10^{-7} $& $ 2.39047\times 10^{-13} $ & $ 2.39039\times 10^{-13} $\\
    & $ 10^{12} $ & $ 8.77694\times 10^{-8} $& $ 2.39021\times 10^{-15} $ & $ 2.39018\times 10^{-15} $\\
    & $ 10^{13} $ & $ 8.80176\times 10^{-9} $& $ 2.39764\times 10^{-17} $ & $ 2.39759\times 10^{-17} $\\
    & $ 10^{14} $ & $ 1.19772\times 10^{-9} $& $ 3.07174\times 10^{-19} $ & $ 3.07167\times 10^{-19} $\\  
\end{tabular}
\end{ruledtabular}
\end{table}
\normalsize
Results presented in Table \ref{tab2} aimed to perform several comparisons, by taking the CME and the laser frequency as $\sqrt{s}=M_{Z}$ and $\hbar\omega=0.117\,eV$, respectively. First, we have compared, for different laser amplitudes, the TCS given by Eq.(\ref{48}) with that given by the Breit-Winger formula (Eq.(\ref{51})), in which the dressing of both incident and scattered particles are taken into consideration. As we can see in the Table \ref{tab2}, regardless of the number of exchanged photons, the two equations lead, nearly, to the same values of the TCS. In addition, this result proves that obtained in Sec.\,\ref{section 2}, which indicates that the Breit-Wigner formula is still valid in the presence of a CP-laser field. Secondly, we have compared the TCS's values described by Eq.(\ref{26}) with those obtained by using Eq.(\ref{48}), in order to show the difference between results of Sec.\,\ref{section 2} and \ref{section 3}. We have concluded from Table \ref{tab1}, that the CP-laser field do not affect the TCS when its amplitude is included in the interval [$10\,V.cm^{-1}$, $10^{4}\,V.cm^{-1}$]. For the laser field's strengths between $10^{5},V.cm^{-1}$ and $10^{8}\,V.cm^{-1}$, as mentioned in Table \ref{tab2}, the muons dressing doesn't show any effect at all on the TCS. Moreover, as long as the laser field's strength overcomes the $10^{8}\,V.cm^{-1}$, the TCS moves from $2.19751\times 10^{-5} [nb]$ to $ 1.52064\times 10^{-10} [nb]$. In addition, the higher the laser field's strength, the bigger the difference between the tow total cross sections. This result is interpreted by the fact that muon's effective mass doesn't change unless the laser field's strength reaches $10^{9}\,V.cm^{-1}$. For example, the effective mass of muon is equal to its corresponding free mass $(m_{\mu}=m_{\mu}^{*}=0.106\,GeV)$ for the laser's electric field amplitudes which are between $\varepsilon_{0}=10\,V.cm^{-1} $ and $\varepsilon_{0}=10^{9}\,V.cm^{-1} $. While for the laser's electric field amplitudes $10^{9}\,V.cm^{-1} $, $10^{10}V.cm^{-1} $ and $10^{11}\,V.cm^{-1} $, the muon's effective mass is $0.106009\,GeV$, $0.106013\,GeV$ and $0.107333\,GeV$, respectively. Therefore, the effective mass progressively increases as long as the laser's electric field amplitude increases. The reader may ask why the laser's effect on electron and muon begins at different laser field's strengths, which are $10^{5}\,V.cm^{-1}$ and $10^{9}\,V.cm^{-1}$, respectively. The answer is that muon's mass is about $200$ times bigger than the electron's mass.  Thirdly, by comparing the results found for the numbers of exchanged photons $-50 \leq  n,N \leq + 50 $ and $-200 \leq  n,N \leq + 200 $, we observe that, for the laser field's frequency $\hbar\omega=0.117\,eV$, the decrease in the TCS is inversely  proportional to the increase in the CP-laser field strength $(\varepsilon_{0})$. Furthermore, for a given laser field's strength, the TCS increases by increasing the number of exchanged photons. For instance, for $\varepsilon_{0}=10^{11}\,V.cm^{-1} $, the TCS is $1.51432\times 10^{-14}[nb]$ and $ 2.39047\times 10^{-13}[nb] $ for the numbers of exchanged photons $-50 \leq  n,N \leq + 50 $ and $-200 \leq  n,N \leq + 200 $, respectively.
\section{Conclusion}
In the present paper, we have investigated the weak process ${e}^{+} {e}^{-} \rightarrow  {\mu}^{+} {\mu}^{-}$ in the presence of a circularly polarized laser field. We have derived the TCS by using both the S-matrix formalism and the Breit-Wigner approach. We have found that the two methods are in excellent agreement as they lead to the same results not only in the case of dressing incident particles but also by dressing both incident and scattered particles. We have also explored the variation of the cross section of the $n$-photon processes as a function of the centre-of-mass energy for different number of exchanged photons. We have found that the total cross section decreases as much as the number of transferred photons decreases and once we reach the corresponding cutoff, the laser's field doesn't show its effect anymore.  The variation of the TCS as a function of the laser field's strength $\varepsilon_{0}$ indicates that the effect of the laser field begins at small intensities for low exchanged photons number and low frequencies. The dependence of the laser-assisted differential cross section on the centre of mass energy is shown for different scattering angles. In Sec.\,\ref{section 3}, we have illustrated the effect of dressing both incident and scattered particles on the total cross section. We have found that this effect begins to appear at different laser field's strengths which are successively $10^{5}\,V.cm^{-1}$ and $10^{9}\,V.cm^{-1}$. This result is interpreted by the fact that heavy particles require high laser field's strength in order to acquire effective masses inside the electromagnetic field.

\end{document}